\documentstyle[draft,psfig]{mn}

\title[The Large Magellanic Cloud cluster NGC 2214]
      {The Large Magellanic Cloud cluster NGC 2214}

\author[T. Banks, R.J. Dodd and D.J.Sullivan]
       {Timothy Banks,$^{1}$ R.J. Dodd$^{2}$ and D.J. Sullivan$^{1}$\\
        $^{1}$Department of Physics, Victoria University of Wellington,
        PO Box 600, Wellington, New Zealand \\
        $^{2}$Carter Observatory, PO Box 2909, Wellington, New Zealand }

\date{Accepted 1995 January 29. Received 1995 January 17;
	in original form 1994 September 12}

\pagerange{\pageref{firstpage}--\pageref{lastpage}}

\def\LaTeX{L\kern-.36em\raise.3ex\hbox{a}\kern-.15em
    T\kern-.1667em\lower.7ex\hbox{E}\kern-.125emX}

\begin{document}
\label{firstpage}
\maketitle



\tolerance=10000

\def\diameter{{\ifmmode\mathchoice
{\ooalign{\hfil\hbox{$\displaystyle/$}\hfil\crcr
{\hbox{$\displaystyle\mathchar"20D$}}}}
{\ooalign{\hfil\hbox{$\textstyle/$}\hfil\crcr
{\hbox{$\textstyle\mathchar"20D$}}}}
{\ooalign{\hfil\hbox{$\scriptstyle/$}\hfil\crcr
{\hbox{$\scriptstyle\mathchar"20D$}}}}
{\ooalign{\hfil\hbox{$\scriptscriptstyle/$}\hfil\crcr
{\hbox{$\scriptscriptstyle\mathchar"20D$}}}}
\else{\ooalign{\hfil/\hfil\crcr\mathhexbox20D}}%
\fi}}

\def\sun{\hbox{$\odot$}}
\def\degr{\hbox{$^\circ$}}
\def\arcmin{\hbox{$^\prime$}}
\def\arcsec{\hbox{$^{\prime\prime}$}}
\def\utw{\smash{\rlap{\lower5pt\hbox{$\sim$}}}}
\def\udtw{\smash{\rlap{\lower6pt\hbox{$\approx$}}}}
\def\fd{\hbox{$.\!\!^{\rm d}$}}
\def\fh{\hbox{$.\!\!^{\rm h}$}}
\def\fm{\hbox{$.\!\!^{\rm m}$}}
\def\fs{\hbox{$.\!\!^{\rm s}$}}
\def\fdg{\hbox{$.\!\!^\circ$}}
\def\farcm{\hbox{$.\mkern-4mu^\prime$}}
\def\farcs{\hbox{$.\!\!^{\prime\prime}$}}
\def\fp{\hbox{$.\!\!^{\scriptscriptstyle\rm p}$}}

\newcommand{\Stromgren}{Str{\"{o}}mgren}
\newcommand{\Grosbol}{Gr{\o}sbol}

\def\loa{\mathrel{\mathchoice {\vcenter{\offinterlineskip\halign{\hfil
$\displaystyle##$\hfil\cr<\cr\approx\cr}}}
{\vcenter{\offinterlineskip\halign{\hfil$\textstyle##$\hfil\cr
<\cr\approx\cr}}}
{\vcenter{\offinterlineskip\halign{\hfil$\scriptstyle##$\hfil\cr
<\cr\approx\cr}}}
{\vcenter{\offinterlineskip\halign{\hfil$\scriptscriptstyle##$\hfil\cr
<\cr\approx\cr}}}}}


	\psfull


\begin{abstract}
Johnson {\it BV} CCD observations have been made of the young Large
Magellanic Cloud cluster NGC~2214 and a nearby field using the
Anglo-Australian Telescope.  It has been suggested in the literature that
this elliptical cluster is actually two clusters in the process of
merging. No evidence is found from profile fitting or the colour--magnitude
diagrams to support this contention.  Completeness factors are estimated
for the CCD frames. These values are used in conjunction with luminosity
functions to estimate the initial mass function (IMF) for NGC~2214. A power
law is assumed for the IMF, with a good fit being found for the exponent (1
+ $x$) = 2.01~$\pm$ 0.09. There is some indication that the low-mass end
($\loa \rm 3\: M_{\sun}$) has a lower gradient than the high-mass end of
the derived IMF.  This value is in reasonable agreement with literature
values for other Magellanic IMFs, and not substantially different from
those of the poorly determined Galactic IMFs, suggesting the possibility of
a `universal' IMF over the Magellanic Clouds and our Galaxy in the mass
range $\sim 1$ to $\sim 10$ $\rm M_{\sun}$.
\end{abstract}



\section{Introduction} NGC~2214 ($ \rm \: \alpha_{2000} \: = \: \rm 6^{h}
\: 12^{m} \: 57^{s}, \delta_{2000} \: = \: 68^{o} \: 15' \: 33'' $ South)
is a young ($ \rm 32 \: \times \: 10^6 $ yr; Elson 1991) populous star
cluster situated in a relatively uncrowded field to the far north-east of
the bar in the Large Magellanic Cloud (LMC).  Meylan~\& Djorgovski (1987)
analysed an intensity profile of the cluster, and found that the core was
abnormal. They conjectured that perhaps it had collapsed, although Elson,
Fall~\& Freeman (1987) have shown that the two-body relaxation time of the
cluster is $\rm \sim2-6 \: \times \: 10^8 $ yr, and so greater than its
age.  Bhatia~\& MacGillivray (1988) found the cluster to have a very
elliptical ($ e \, = \, 0.5$) core with an almost spherical halo, and
suggested that this unusual shape could be due to NGC~2214 being a binary
star cluster in an advanced stage of merging.  Comparison with N-body
simulations lent support to this idea. Sagar, Richtler~\& de Boer (1991a)
used the 1.54-m ESO Danish telescope in $ \sim1$-arcsec seeing, and
presented a {\it BV} colour--magnitude diagram (CMD) with two well-defined
supergiant branches, separated by $ \sim2$ mag in {\it V}. The older
population was more centrally condensed than the younger one, and Sagar~{et
al.} (1991a) suggested that the first published CMD (Robertson 1974) had
failed to detect the older branch due to the problems of photometry in such
a crowded region.

A major objective of the present study was to derive an estimate of the
initial mass function (IMF) of the cluster. The IMF is defined as the
frequency distribution of stellar masses on the main sequence at the
formation time of a group of stars (Scalo 1986).  Mass is one of the
primary factors influencing stellar evolution, and a detailed knowledge of
the IMF would be important in a wide range of studies ranging from galactic
evolution to the spectral properties of binary stars (see Tinsley 1980).  A
fundamental question about the IMF is whether it is universal in time and
location, or whether the distribution of stars formed is a function of
parameters such as metallicity.

Derivation of the IMF is not straightforward. An initial approach might be
to use the nearby solar neighbourhood to do this, but this technique is
complicated by the fact that these stars have a range of distances, ages,
and metallicities.  For instance, the random velocities of the stars,
combined with their lifetimes, means that, while massive stars will still
be near the site of their formation, low-mass stars will have travelled
significant distances. Variations in composition may result just from such
spatial considerations, if not from galactic evolution as well. Scalo
(1986) comments that the many assumptions, such as any variation in the
star formation rate with time, complicate estimates of the field IMF to the
point of impracticality. In addition, a universal nature is assumed for the
IMF in such studies.

A better approach is to use clusters, where the component stars will be
effectively coeval and of the same composition. Such work is complicated by
effects such as dynamical evolution leading to mass segregation in the
cluster, tidal stripping (which in the presence of mass segregation will
lead to the proportional decrease of low-mass stars; see Spitzer 1987), and
stellar evolution as stars evolve off the main sequence, which leads to no
easily derivable mass function information for stars of a main sequence
lifetime less than the age of the cluster. The mass function of the cluster
may alter substantially with time, and it is best to select young clusters
where these effects have not had time to become significant. Many studies
have centred on young Galactic open clusters with their large observable
mass range (e.g. Phelps~\& Janes 1993; Reid 1992; Stauffer~{et al.}
1991). However, such work is complicated by field star contamination,
counting incompleteness, and low number statistics (see Scalo 1986 for more
details), as well as the problem that most open clusters suffer substantial
and variable reddenings due to their positions in the Galactic disc (Mateo
1988). There is no strong evidence for variations in the shapes of their
mass functions (Sagar~\& Richtler 1991). Globular clusters offer better
statistics due to the increased number of stars they contain, but the
observable mass range is limited due to their distances and
age. Evolutionary effects, such as mentioned above, are additional
complications. The resulting mass functions appear to vary considerably
between clusters, and may be correlated with metallicity (Sagar~\& Richtler
1991), although this is clouded by the above problems.

The LMC clusters are effectively a mixture of the best features of these
two types of star clusters. They are populous, with resultingly good
statistics, and span a wide range of ages and metallicities (Da Costa
1991). The clusters are distant enough to subtend only a small angle on the
sky, and yet not too distant to suffer from resolution problems. Questions,
such as the universal nature of the IMF, might be able to be addressed
using these clusters, although the very populous nature of both the
clusters and their fields leads to counting incompleteness problems. A
major portion of this study involved the derivation of counting estimates,
in order to correct observed luminosity functions to the `real'
distribution.

IMFs have been derived for some LMC clusters by Mateo (1988), Sagar~\&
Richtler (1991), Cayrel, Tarrab~\& Richtler (1988), and Elson, Fall~\&
Freeman (1989). The results have not been in good agreement. The first three
studies were based on CCD frames, and attempted to estimate the counting
incompleteness using artificial star trials (see below). A power law ${
\frac{{\rm d}N}{{\rm d}M} \: = \: M}^{-(1 \: + \: x )}$ was assumed for the
IMF, where d$N$ is the number of stars in a given mass interval d$M$ at
mass $ M . $ Mateo (1988) found that the IMFs of six Magellanic clusters (the
Small Magellanic Cloud cluster NGC 330 was included) could all be fitted
with the single power law with $x$ = $ 2.52 \: \pm \: 0.16$ over the mass
interval 0.9 to 10.5 $\rm M_{\sun}$.  Sagar~\& Richtler (1991) used a
different method of estimating the incompleteness (see below), and arrived
at an $x$ value of $\sim$1.1, not too different from the Salpeter (1955)
value of 1.35 and in reasonable agreement with the value of 1.2 for NGC 330
and NGC 1818 derived by Cayrel~{et al.} (1988). They commented that if they
used the same incompleteness technique as Mateo (1988) on NGC 1711, which
was the only cluster studied by both, then the mass function estimate of
Mateo (1988) was confirmed. All these values contrast sharply with the
photographic star count analysis of Elson~{et al.} (1989), which gave $x$
values between $-0.2$ and 0.8 (over 1.5--6.0 $\rm M_{\sun}$).  In light of
these differences and the comment of Sagar~\& Richtler (1991) about NGC
1711, a review of the incompleteness techniques is obviously of major
importance given the effect a chosen method has on the derivation of the
mass function slope, and any subsequent conclusions about the universality
of the IMF.


\section{Observations}

Johnson {\it BV} observations of NGC~2214 were collected on the night of
1993 March 1/2 using a 1024 by 1050 pixel TEK CCD at the prime focus of the
Anglo-Australian Telescope. The pixel scale was 0.39~arcsec per pixel,
resulting a field of view approximately 6.7 $\times$ 6.7 arcmin square.  The
FWHM seeing was $ \sim2.2 $-arcsec.  Observations were also made of a field
5-arcmin north of the cluster.  Exposure times for both these regions were
30 and 300~s in {\it V}, and 60 and 600~s in {\it B}.

The initial reductions of the CCD frames were carried out at the
Anglo-Australian Observatory, and included trimming the frames of the
overscan rows, subtraction of the mean of the overscan (the CCD has
negligible bias structure), and flat-field division using sky flats. Given
already derived extinction coefficients (Da Costa, private communication),
the observed Graham~(1982) E2 and E3 standards were used in the {\sc iraf}
{\sc photcal} package to derive the zero-point shift in the following
transformation equations:

\[
b  =  ( 1.067  \! \pm \! 0.007 ) \! +  \! 0.12 ( B \! - \! V ) \! - \!
\left( 0.4 \!  - \! 0.02 ( B \!  - \! V ) \right) X \! + \! B
\]
\[
v  =  ( 1.093  \pm  0.006 )  -  0.27  X  + V
\]
where $X$ is the airmass and the lower case letters refer to the observed
instrumental magnitudes. The root mean squares of the fits were 0.018 and
0.016 mag for {\it B} and {\it V} respectively.


\section{Results}

\subsection{Colour--magnitude diagrams}

\begin{figure*}
{
\centerline
	{
	\psfig{figure=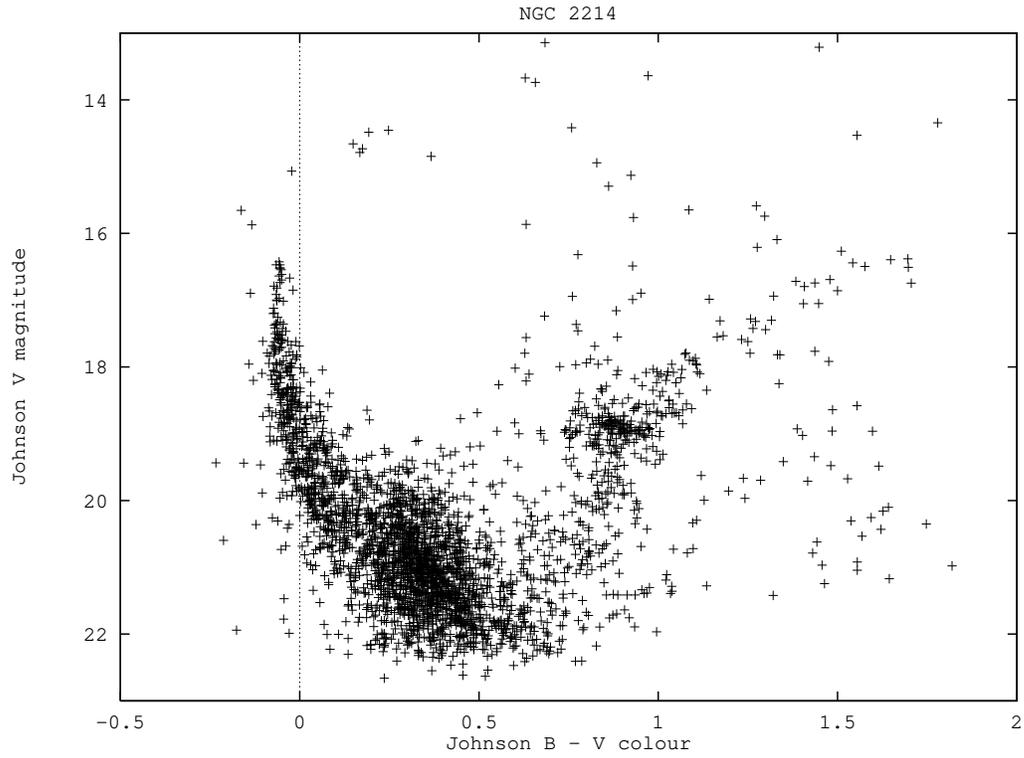,height=10cm,width=14cm,angle=-90}
	}
\caption[Colour--magnitude diagram for NGC~2214]{{  Colour--magnitude diagram
for NGC~2214.}
\label{figure:mot} }
}
\end{figure*}

\begin{figure*}
{
\centerline
	{
	\psfig{figure=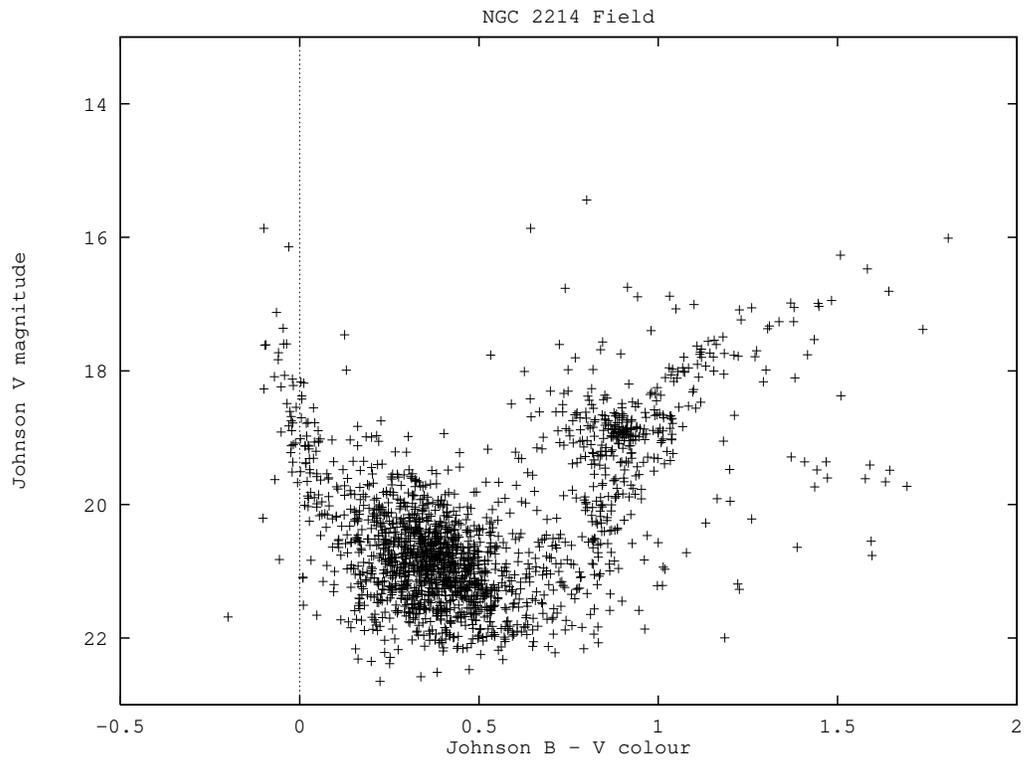,height=10cm,width=14cm,angle=-90}
	}
\caption[Colour--magnitude diagram for the Field]{{  Colour--magnitude diagram
for the Field.}
\label{figure:mot2} }
}
\end{figure*}

The {\sc digiphot} package of {\sc iraf}, which includes {\sc DAOphot} (Stetson
1987), was used to reduce the crowded frames, perform aperture corrections,
and transform the data across to the standard system. The final
colour--magnitude diagrams are given as Figs~\ref{figure:mot} and
\ref{figure:mot2}. 2919 and 1832 stars are plotted in the cluster and field
diagrams respectively. The matching point between the long and short
exposures was chosen to be the region where the data sets had similar
errors, and was magnitudes 16 and 17 for {\it V} and {\it B} respectively.

\label{uno}

\begin{figure*}
\centerline
	{
	\psfig{figure=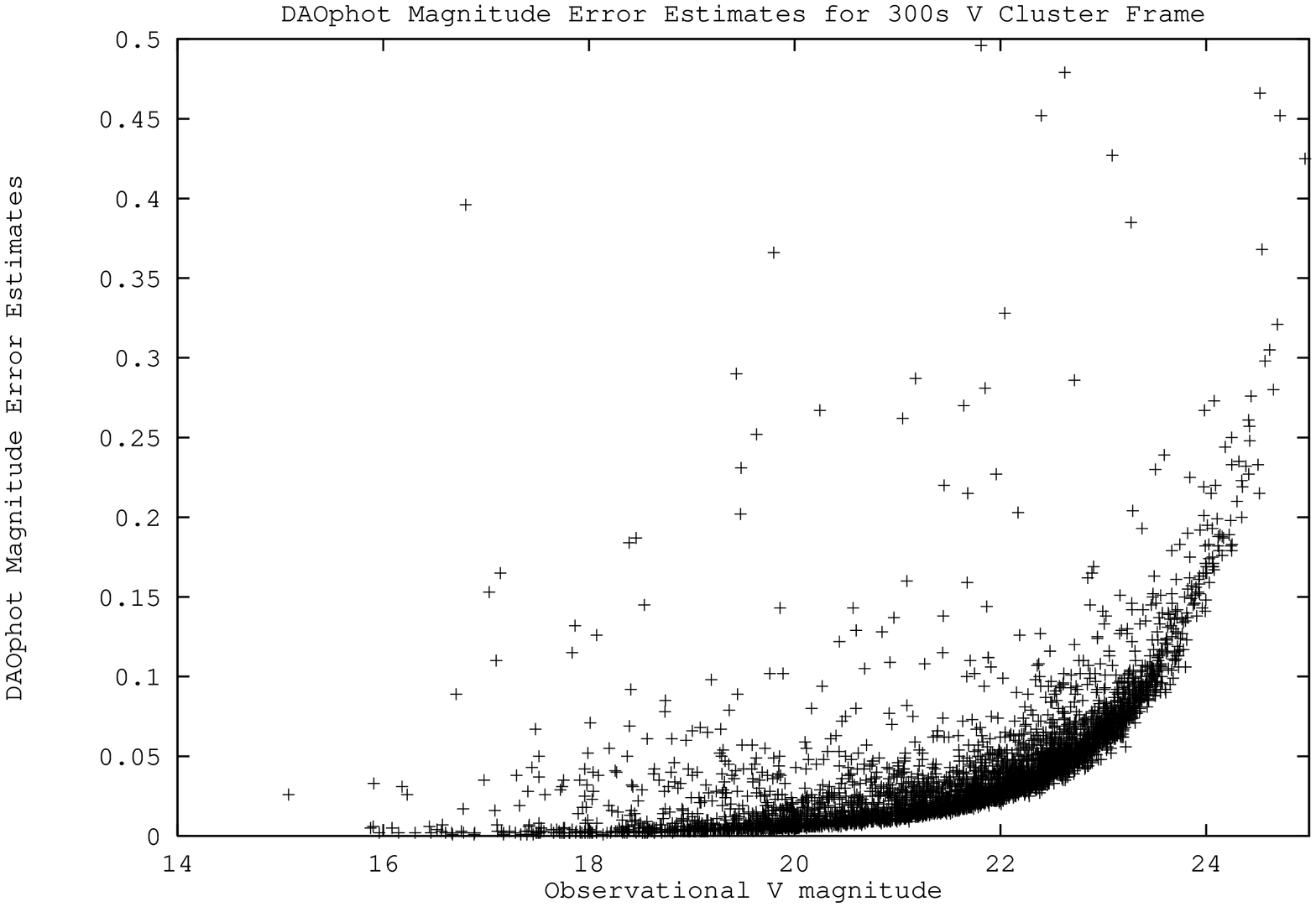,height=9.5cm,width=14cm,angle=-90}
	}
 {\caption[Photometric errors for the 300-s {\it V} exposure of NGC~2214]
{{  Photometric Errors for the 300 second {\it V} exposure of NGC~2214.}
\label{figure:hai} } }
\end{figure*}

$\chi$ is the ratio of the actual scatter, about a point spread function
(PSF) fit, divided by the expected scatter given the star and background
sky brightnesses combined with the CCD readout noise. A value of $\chi$
near unity indicates a good fit. Only stars with a $\chi$ value of 3.0 or
less were accepted (as in Mateo~\& Hodge 1986). Examination of $\chi$
plotted against observational magnitude showed this to be an acceptable
limit, with the vast majority of detected `stars' being within it.  A
further constraint was the use of the `sharpness' measure of the difference
between the square of the width of the object and the square of the width
of the PSF.  Values should be close to zero for single stars, large and
positive for blended doubles and partially resolved galaxies, and large and
negative for cosmic rays and blemishes. Examination of this parameter for
all detected `stars' showed that the majority had values inside
$|0.2|$. The final selection criterion was the use of photometric error
estimates from DAOphot. Photometric errors, as determined from least-squares
fitting by DAOphot, naturally increase with magnitude (see
Figs~\ref{figure:hai}). If a low threshold were set for the acceptable
errors across all the data, the fainter magnitudes would be excluded. In
order to retain as many as possible of these fainter stars in the CMD, the
acceptable photometric uncertainty was relaxed as magnitude
increased. These values are given in Table~\ref{table:one}.

\begin{table}
\caption[Adopted error limits for NGC Cluster and Field Photometry]{ { Adopted
error limits.} \label{table:one} The majority of the photometric errors derived
by DAOphot fall on to an exponential function. The error adopted for each
observational magnitude range (as given in the table) corresponds to the value
of this function at the fainter magnitude limit of each interval. The number of
stars recovered in each bin is also indicated, under the headings `Cluster' and
`Field' for each filter. }
\scriptsize{{\renewcommand{\tabcolsep}{5pt}
\begin{center}
\begin {tabular}{@{}c|c|r|r||c|c|r|r@{}}
\hline
\multicolumn{4}{||c||}{Johnson $ V $} & \multicolumn{4}{c||}{Johnson $ B $} \\
\cline{1-8}
Mag & Limit & Cluster & Field & Mag & Limit & Cluster & Field \\
\hline
10-16 & 0.010 & 16 & 1 & 10-17 & 0.001 & 22 & 2 \\
16-18 & 0.011 & 56 & 13 & 17-18 & 0.005 & 33 & 5 \\
18-19 & 0.016 & 111 & 54 & 18-19 & 0.013 & 109 & 22 \\
19-20 & 0.025 & 349 & 139 & 19-20 & 0.018 & 236 & 103 \\
20-21 & 0.040 & 504 & 291 & 20-21 & 0.028 & 550 & 317 \\
21-22 & 0.060 & 925 & 644 & 21-22 & 0.040 & 886 & 578\\
22-23 & 0.090 & 1045 & 811 & 22-23 & 0.063 & 1055 & 873 \\
23-24 & 0.200 & 287 & 207 & 23-24 & 0.120 & 568 & 349 \\
\hline
Total: & - & 3293 & 2160 & Total: & - & 3459 & 2249 \\
\hline
\end{tabular}
\end{center}
}}
\end{table}

The main features of the cluster CMD are the following.
\begin{enumerate}
\item A conspicuous main sequence of ($B - V$)
$ \sim $ 0 extending to {\it V} $ \sim $ 16.5.
\label{enumerate:lone}
\item Evolved stars above this main sequence turnoff.
\label{enumerate:ltwo}
\item A subgiant branch of ($B - V$) $ \sim $ 0.85,
extending from {\it V} $ \sim $ 19 to $ \sim $ 21.
\item A giant star clump at ($ V $, $B - V$) $ \sim $ (19, 0.9)
which is slightly extended in
{\it (B-V)} by some 0.25. This extension is above
the accepted photometric error at this
{\it V} magnitude (see Table~\ref{table:one}).
\item A uniformly populated giant branch extending from
the giant clump up to ($ V$, $B - V$)
$ \sim $ (16.5, 1.7).
\item Several stars blueward of the main sequence,
which are artefacts of the crowded field reduction (Lee 1992),
being stars in dense regions whose parameters could not be estimated
accurately by DAOphot. Magnitudes tend to be under-estimated.
This is the basis of the bin migration phenomena noted by Mateo (1988),
amongst others.
\item A clump peaking at ($ V $, $B - V$) $\sim$ (19.7, 0.3) of older stars.
\item Many faint red stars in the lower right of the CMD, which have been
tentatively identified as giant stars by Chiosi (1989),
and considered to be field stars.
\end{enumerate}
Only items \ref{enumerate:lone} and \ref{enumerate:ltwo} are features
associated with the cluster, the others being due to field stars. Being at
a moderate Galactic latitude ($\sim \rm -30^o $) the majority of field
stars will be in the LMC. The field CMD appears to be contaminated to an
extent by cluster stars, as evidenced by the reasonably strong main
sequence.

As mentioned in the Introduction, Sagar et al. (1991a) presented a CMD for
NGC~2214 with two supergiant branches, and claimed that, since the
population of the older branch was more centrally concentrated than that of
the younger one, previous studies would have missed them due to the
increased level of crowding. The shallow {\it BV} CMD of Elson (1991)
showed no sign of the second turnoff, although the $ \sim5.5$-arcsec seeing
may have obscured it, as could the 3-arcsec seeing in the {\it BV} CMD of
Banks (1993).  However, examination of colour--magnitude diagrams for other
LMC clusters presented in Sagar, Richtler~\& de Boer (1991b) showed traces
of second (weak) supergiant branches in other clusters. This, combined with
the high number of stars blueward of the cluster main sequence, which are
an indicator of crowded field reduction problems (Lee 1992), suggested that
there may be problems with the reduction process.  Short-exposure
observations taken with the Mount John University Observatory (NZ) 1-m
telescope gave some indication of a clump in the main sequence at the {\it
V} magnitude corresponding to the older turnoff (Banks 1993; Banks, Dodd~\&
Sullivan 1994).  After submission of the observing proposal for the current
study, which involved an investigation of the reality of the second branch,
Lee (1992) published a paper describing CCD observations of NGC~2214 in
1.1-arcsec to 1.6-arcsec seeing with the Las Campanas du Pont 2.5-m
telescope.  The resulting {\it BV} CMD showed only one supergiant branch.
The main sequence was matched well by a Maeder~\& Meynet (1991) isochrone
for 50-Myr, while the supergiant branch was approximately matched by an
older 70-Myr isochrone.  The current study provides no evidence for a
second supergiant branch.

CMDs were also generated for subsections of the frame, in order to search
for differences across the cluster which might be expected for two merging
clusters as in Bhatia~\& MacGillivray (1988). First the frame was split
vertically about the cluster centre, which was estimated as the mean value
derived from the ellipse fitting described below. The rotation of the
cluster was also estimated as the mean value. Marginal distributions peak
close by.  Given that the position angle was $\sim100$\degr\ , this
division effectively split NGC~2214 into the suggested merging halves.  No
difference was found in the shapes of the two CMDs, nor those for four
300~$\times$ 300 pixel regions placed at 90\degr increments about the
cluster centre, suggesting that if the merger idea is correct then the two
bodies are of similar age (see also Bhatia~\& Piotto 1993, who came to the
same conclusion).


\begin{figure}
\centerline
	{
\psfig{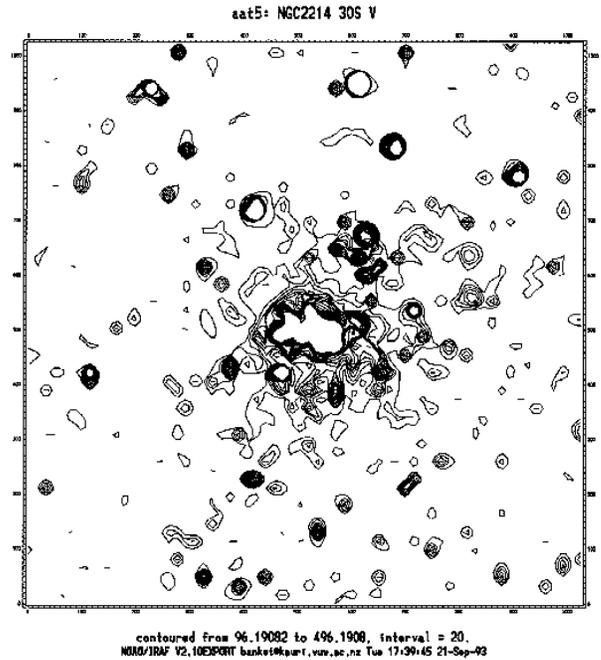}
	}
\caption[Contour Maps of the 30s NGC~2214 {\it V} frame]
{{ Contours for the outer regions of the 30-s exposure {\it V} frame of
NGC~2214. The contour lines
are increments of 20 counts per pixel over the range 96 to 496. Contours
for the inner regions are shown in Figure~\ref{figure:freddy}.}
\label{figure:contours} }
\end{figure}

\begin{figure}
\centerline
	{
\psfig{figure=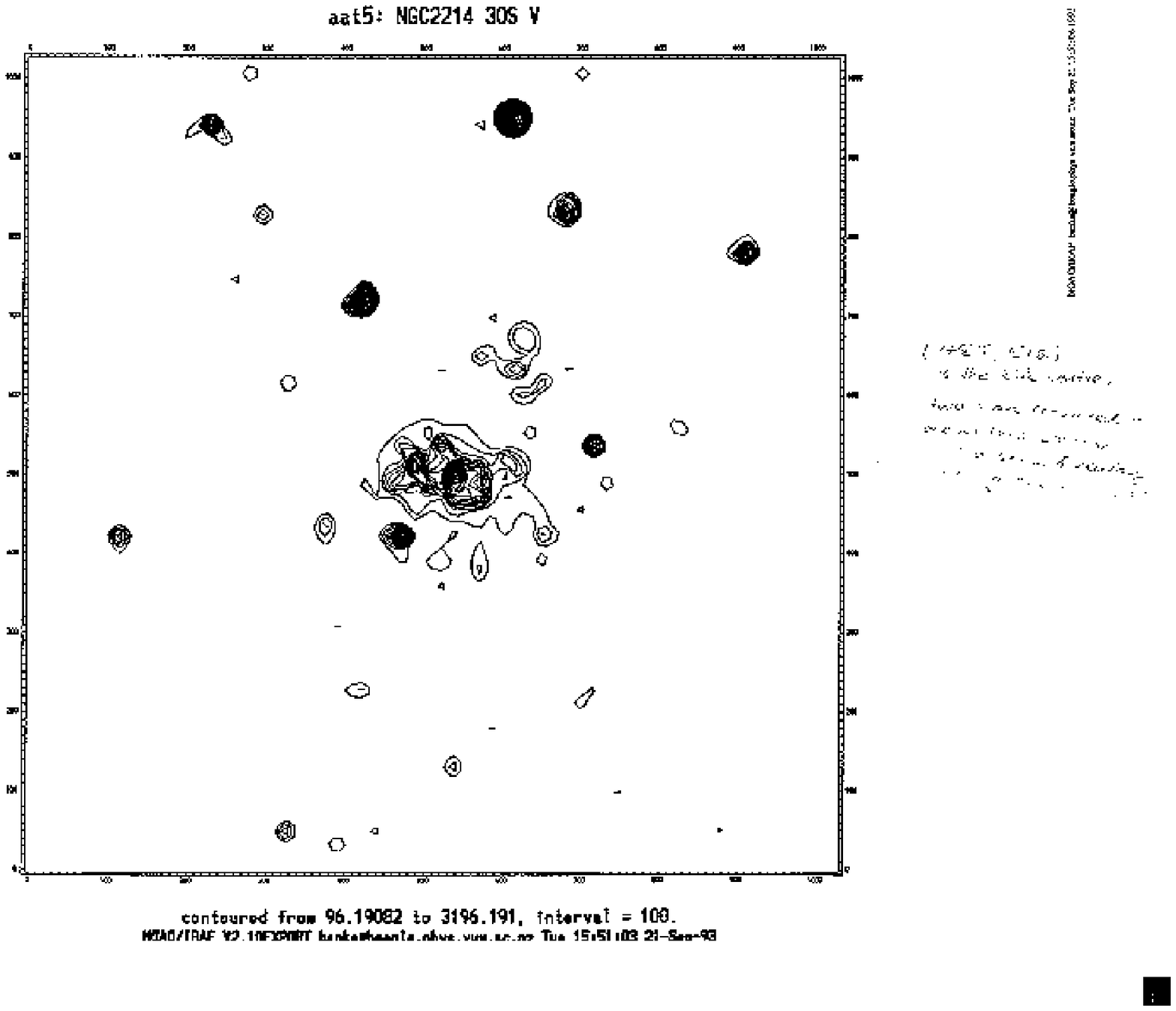,height=9.5cm,width=9cm,bbllx=38pt,bblly=218pt,bburx=451pt,bbury=640pt,clip=t}
	}
\caption{Contours for the inner regions of NGC~2214, in which contour
levels range from 196 to 3196, with an increment of 100. Note the apparent
double nature of
the core, which should be compared with Fig.~2b of Bhatia~\& MacGillivray
(1988).
\label{figure:freddy}}
\end{figure}


\begin{figure}
\centerline
	{
	\psfig{figure=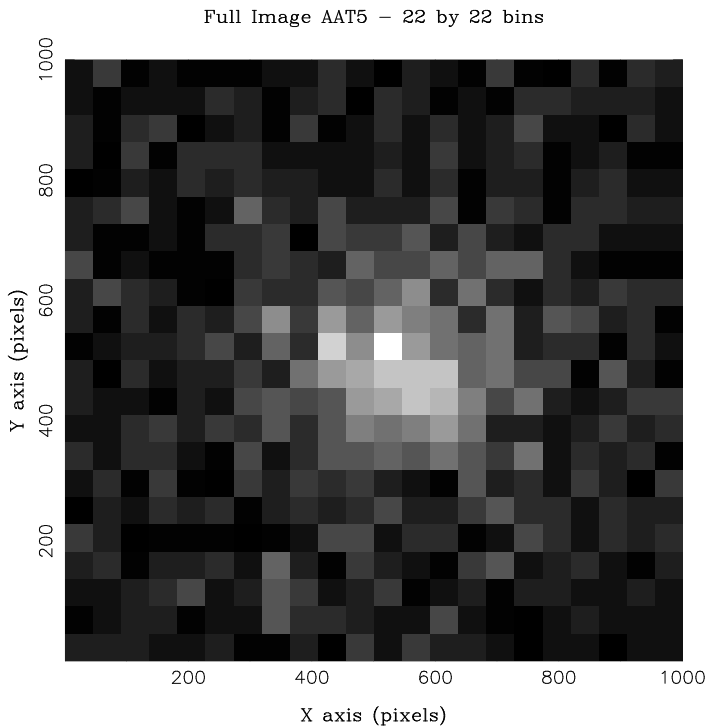,height=10cm,width=13.8cm,angle=-90}
	}
{ \caption[NGC~2214 30-s {\it V} frame star counts in 19.5-arcsec bins]
{{  30-s {\it V} cluster frame star counts in 19.5-arcsec bins}. White
corresponds
to the greatest number of stars (220 or more per square arcminute), while black
corresponds to 15 or less
stars per square arcminute. \label{figure:grey1}}}
\end{figure}

\begin{figure}
\centerline
	{
\psfig{figure=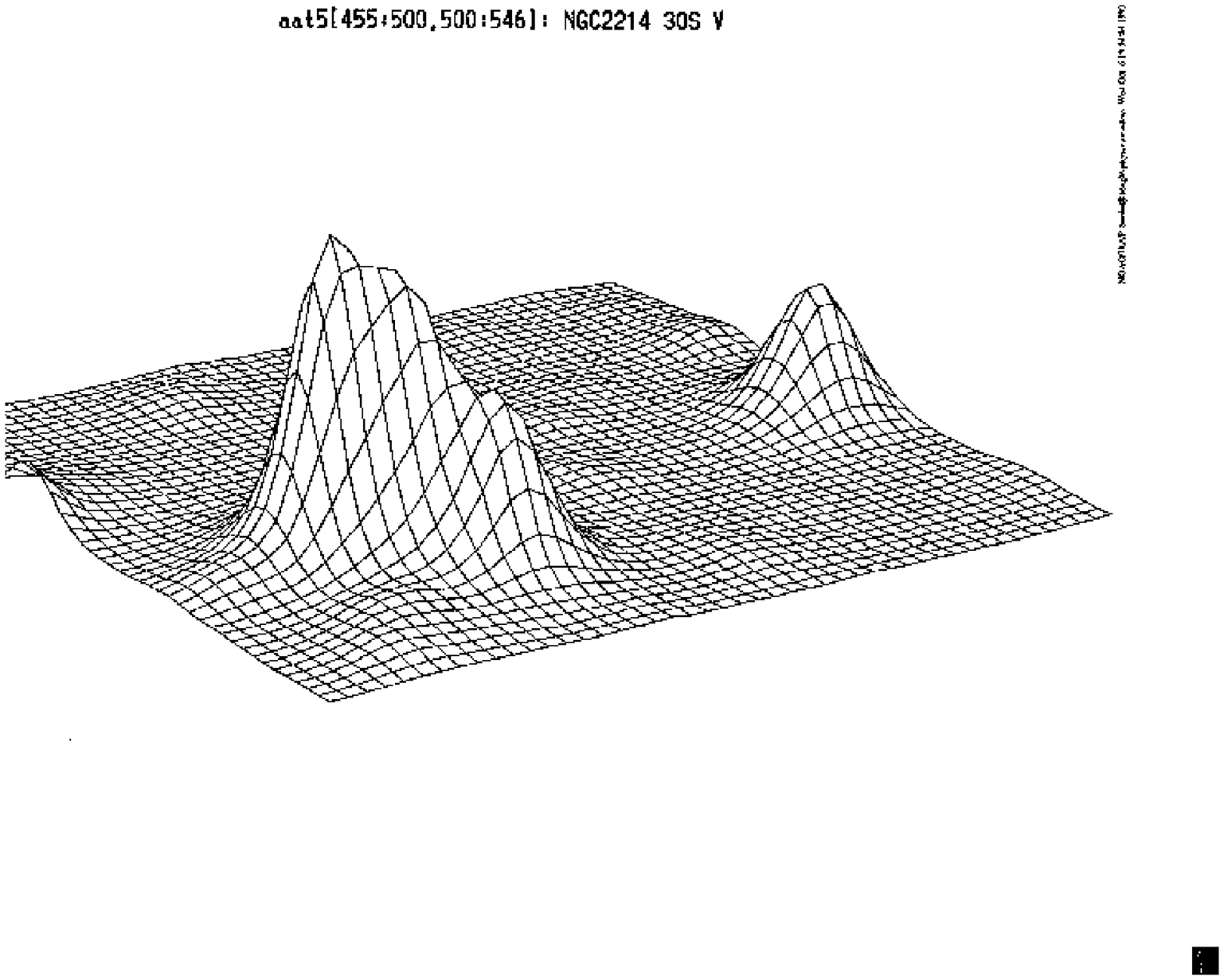,height=4cm,width=8cm,bbllx=38pt,bblly=280pt,bburx=500pt,bbury=510pt,clip=t}
	}
\caption[Grid diagram of the blended region]{{  Grid diagram of the blended
region.}
A surface diagram, or a three-dimensional
representation of a portion of the frame with intensity as the
{\it z} (vertical) axis, shows that the gap between the two  `cores'
\label{figure:grid_diagrams} contains the blended product of
several bright stars. Each grid square
represents one pixel.
}
\end{figure}

\subsection{Contours}

Bhatia~\& MacGillivray (1988) presented contours of NGC~2214 based on a
IIIaF UK Schmidt survey plate, which exhibited two lobes in the core of the
cluster (see also Figs~\ref{figure:contours} and \ref{figure:freddy}).
Noting the possible criticism that isophotal maps may be
influenced by the placement of bright stars, Bhatia~\& MacGillivray (1988)
processed the COSMOS (MacGillivray~\& Stobie 1984) scans of the plates
using a crowded field package in order to obtain star counts. They claimed
these would not be strongly influenced by bright stars and so reflect more
appropriately the distribution of stars in the cluster. Again a double peak
was evident, although it can be seen that the contour delineating the peaks
is not statistically different from the next `lower' contour, which does
not show lobes, if a Poisson distribution is assumed.  It should also be
noted that the IIIaJ plate contours presented by Bhatia~\& MacGillivray
(1988) do not show a double peak, but rather show a slightly offset
centre. This suggests that the lobes are due to a small number of stars,
with markedly different colour indices.

The results of contouring the 30-s {\it V} frame of NGC~2214 are shown in
Figure~\ref{figure:contours}.  As in Bhatia~\& MacGillivray (1988), two
components appear to be in the cluster centre.  Following Bhatia~\&
MacGillivray (1988), stars were counted into 0.3~$\times$ 0.3 $\rm
arcmin^{2}$ bins, and each scaled to the number of stars per
arcminute. A central condensation is evident, with a slight indication of a
lobe, although a grey-scale representation of the data shows that the
contour diagram tends to over-interpret the data (see
Fig.~\ref{figure:grey1}). The central peak bin contained 20 stars, while
the alleged secondary peak and gap between the two peaks contained 15 and
11 stars respectively. Assuming a Poissonian distribution, these latter two
values are not statistically different.

Different selection effects apply to each of the bins, and so like is not
being compared with like.  The gap contains three of the four brightest
stars in the bins.  These stars, and the blended product shown in
Fig.~\ref{figure:grid_diagrams}, could be obscuring fainter stars. This
would explain why the faintest star recovered in this bin is 2.5~mag
brighter than that recovered in the less crowded `second lobe'. Certainly
only the two outer stars were detected in the blended peak, with another
further possible star being lost in the outer regions of the brightest
star. The latter `star' appears as a small rise in the outer regions of the
point spread function (PSF) of the bright star when it is compared with the
generic PSF derived for the frame.

The second peak is mainly outside the populated central region of the
cluster, where blending is less severe.  Fainter stars can be resolved, and
so the star count increased. Star counts in populous clusters cannot
naively be considered to reflect the underlying stellar distribution, due to
problems deconvolving extremely crowded regions. A worst case scenario for
a populous spherical cluster could show only a few star counts in the
heavily blended inner regions, surrounded by a ring of greatest
counts. From the centre of such a cluster, star counts would increase with
radius until the radius of this ring was reached.  Beyond this radius, the
counts would drop away again.

An increase in the number of bins along each axis of the frame resulted in
grey-scale plots of greater resolution, but with the tradeoff of less
counts and greater noise. None of these plots showed any substantial
evidence of a second component to the core.  We contend that there is no
evidence from star counts to support the notion that two cores are present
in this cluster.


\subsection{Ellipse fitting}

\begin{figure}
\centerline
	{
	\psfig{figure=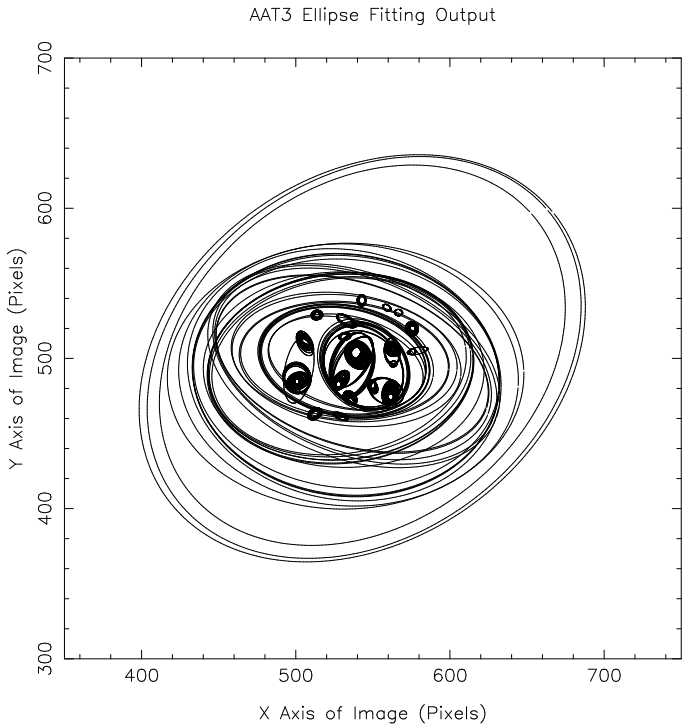,height=10cm,width=13.8cm,angle=-90}
	}
\caption[Ellipse fits to the 60-s {\it B} frame]
{ {  Ellipse fits to the 60-s {\it B} frame.} The `lowest' threshold used
to detect pixels for ellipse fitting was 3 times the standard deviation
of the background (which was in turn found
by using the {\sc iraf} {\sc imex} task in 21 regions across the frame).
\label{figure:ba} }
\end{figure}

Moment-based ellipse fitting was performed on the 60-s {\it B} frame of
NGC~2214 using the routines described in Banks, Dodd~\& Sullivan (1995).  A
generally constant ellipticity of 0.4 was found out to a radius of $
\sim30$-arcsec from the cluster centre (see Fig.~\ref{figure:ba}),
before falling away to an ellipticity of 0.2 by $ \sim40 $-arcsec. Beyond
this radius the ellipticity is constant (out to the maximum measurement of
62-arcsec). The position angle was constant at $\sim100$\degr\ out to a
radius of $\sim$40-arcsec, and then climbed to another plateau at
$\sim150$\degr. This may have been caused by the lower detection
thresholds, used for the greater radii, encountering a grouping of a few
stars to the side of the main body of the cluster, and so skewing the
ellipse fit.  These results agree well with the ellipse fitting of {\it V}
CCD observations of NGC~2214 acquired at the Mount John University
Observatory (NZ), which found similar trends and values even when the
brightest stars were subtracted from the frame.

Zepka~\& Dottori (1987) fitted ellipses to photographic observations of
NGC~2214, and found that the cluster ellipticity was constant at around 0.4
out to their maximum radius of 43-arcsec, and that the position angle of
the clusters was constant.  Frenk~\& Fall (1982) gave an ellipticity of
0.29 for their eye measurements of the burnt-out centre of NGC 2214 in an
SRC Sky Survey plate. This result is in accord with those of the current
study, as the measurement would correspond to outer regions of the cluster.

However, these apparent trends should be treated with caution, given the
results of Banks~{et al.} (1995) in which synthetic frames were generated
of clusters with known ellipticities. Spurious trends were found by both
the moment analysis technique used here, and the method of Jedrzejewski
(1987) as implemented in {\sc stsdas}, due to bright stars, some of which
were so badly blended in crowded regions that they could be neither
detected nor cleanly subtracted.

The ellipse fitting provides no definitive support for a double core to the
cluster.

\begin{table}
\caption[NGC~2214 Isochrone fit ages] { { Isochrone fit ages.}
\label{table:schaerer} Logarithm of the ages are given for the best isochrone
fits to these
features in the CMDs. Different metallicities ({\it Z}) were trialed.
A distance modulus of 18.4~mag was assumed. The long and short
moduli of 18.2 and 18.7 mag were also fitted, with no difference in ages for
the
modulus 18.2, while modulus 18.7 ages were log 0.2 younger. In the paper we
have
used the intermediate distance modulus of 18.4.  The Swiss models are those of
Schaerer~{et al.} (1993), Schaller~{ et al.} (1992), and Meynet~{et al.}
(1993). }
\vspace{-3mm}
\scriptsize{{\renewcommand{\tabcolsep}{5pt}
\begin{center}
\begin {tabular}{@{}c|c|c|c||c|c@{}}
\hline
CMD & \multicolumn{3}{c||}{Swiss models $Z=$} &
\multicolumn{2}{c||}{Bertelli~{et al} $Z=$} \\ \cline{2-6}
Feature & 0.001 & 0.008 &  0.020 &  0.004 & 0.020 \\
\hline
Main Sequence: & - & 8.0 & 8.0 & 8.3 & 8.2  \\
Supergiant Branch: & - & 7.8 & 7.8 & 8.0 & 7.9  \\
Red Giant Clump: & 8.7 & 8.7 & 8.6 & - & 9.0 \\
\hline
\end{tabular}
\end{center}
}}
\end{table}


\subsection{Isochrone fits}

Initially both non-overshooting and overshooting isochrones were
constructed with metallicity {\it Z } = 0.001, 0.008, and 0.020
(Schaerer~{et al.}  1993; Schaller { at al.}  1992; Meynet, Mermilliod~\&
Maeder 1993).  Fits to features in both the cluster and field CMDs were
attempted (see Table~{\ref{table:schaerer}). The data points were
transformed for a distance modulus of 18.4 and reddening of 0.08 (see Lee
1992; Castella, Barbero~\& Geyer 1987; Elson 1991). The {\it V} extinction
was taken to be 3.1 times $E$($B - V$) (Rieke~\& Lebofsky 1985; Mateo 1988;
Sagar~\& Richtler 1991).

The 0.001 isochrones did not fit the features of the CMD at all well, being
too blue. The fit to the red giant clump was poor, due to different
gradients.  Similarly the gradient of the giant branch was too
steep. Non-overshooting models, at all three metallicities, did not fit
this feature well either.  However, the overshooting {\it Z}~=~0.020
isochrones did fit the branch well, but not the field giant clump and the
main sequence at the fainter magnitudes.

\begin{figure*}
\centerline
	{
	\psfig{figure=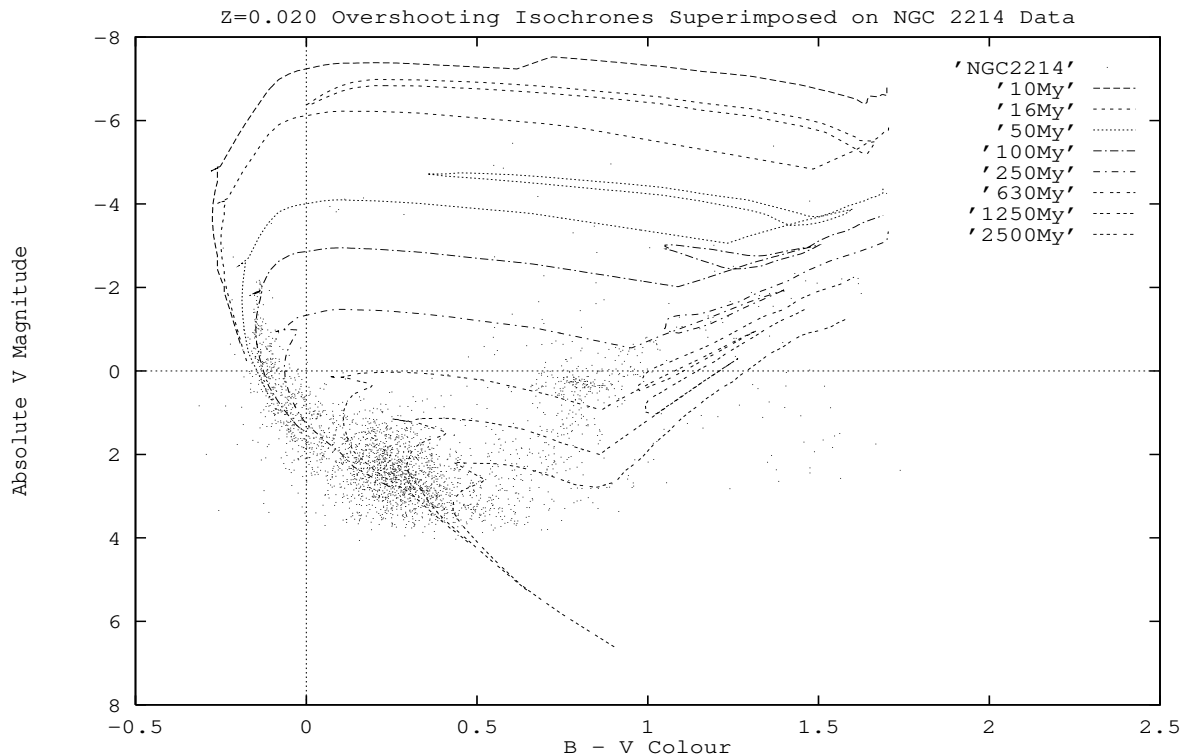,height=10cm,width=16cm,angle=-90}
	}
\caption[{\it Z}~= 0.020 isochrones superimposed on to the {\it BV} CMD of NGC
2214]
{ { {\it Z}~= 0.020 isochrones superimposed on to the {\it BV} CMD of NGC
2214.}
Isochrones based on the Schaller~et al. (1992) evolution models are
plotted over the CMD derived for NGC 2214. The cluster data points have been
shifted for a distance modulus of 18.4 mag and reddening of 0.08 mag (see Lee
1992;
Castella~et al. 1987; Elson 1991). The {\it V} extinction was taken to be 3.1
times E({\it B-V}).
\label{figure:isochronesplot}  }
\end{figure*}

$Z$ = 0.020 isochrones fitted all features of the CMD best, with the {\it
Z}~= 0.008 isochrones slightly too blue.  The ages in
Table~{\ref{table:schaerer} tend to depend more on the {\it V} magnitude
axis than on colour, explaining the similarity of ages derived using the
different metallicities.  It was not possible with any of the isochrones to
have one fit simultaneously the main sequence and the supergiant branch of
the cluster. Lee (1992) used the Maeder~\& Meynet (1991) isochrones, and
encountered the same problem, having to assign different logarithm ages of
7.7 and 7.8 respectively to these features. Sagar~{et al.} (1991a) gave a
range of ages for their `bright' supergiant branch, ranging from the log
value 7.5 using the Castellani, Chieffi~\& Staniero (1990) isochrones, to
the log value 8.0 using Bertelli~{et al.} (1990). The lower age was also
derived by Elson (1991), based on the models of Becker (1981) and
Brunish~\& Truran (1982).  Our results tend towards the upper end of this
age range, since we have used models incorporating convective overshooting,
being, as are the other age estimates, dependent on the model used to
derive the isochrones.

Turning to another set of models, the {\it Z}~= 0.020 isochrones of
Bertelli~{et al.} (1990) fitted the majority of the CMD features well (see
Fig.~\ref{figure:isochronesplot}), confirming the comment of Sagar~{et
al.  } (1991a) that this metallicity is the best fit from the Bertelli ~{et
al.}  (1990) tables of {\it Z} = 0.020, 0.004, and 0.001.\footnote{This is
interesting in light of the Richtler~\& Nelles (1983) \Stromgren\ estimate
of [Fe/H] = $-1.2$~$\pm$ 0.2, which would make NGC 2214 the least
metal-abundant
of all the young LMC clusters (see Da Costa 1991). Richtler~\&
Nelles (1983) also give [Fe/H] values of $-1.6$ and $-1.8$~$\pm$ 0.2 for
the Magellanic Cloud clusters NGC 1818 and 330 (the latter being in the
SMC). Later values for these clusters are $-0.9$ (Richtler, Spite~\& Spite
1989) and $-1.3$ (Spite~{et al.}  1986), both using high-dispersion
spectra, suggesting that Richtler~\& Nelles' (1983) estimate for NGC~2214
is too metal-poor. See also Jasniewicz~\& Th\'{e}venin (1994) who found
values of $-0.37$~$\pm$ 0.03 and $-0.55$~$\pm$ 0.04 for the two clusters.}
Again (see Table~{\ref{table:schaerer}) two isochrones were needed to fit
the upper main sequence and the supergiant branch well, although a shift of
$ \sim $ +0.035 in {\it (B-V)} would allow the 100-Myr isochrone to fit
both.  Such a shift is just outside the formal errors in the combined
aperture corrections and photometric transformations.  The same shift would
allow the {\it Z}~= 0.020 $\rm 6 \: \times \: 10^7 $ yr isochrone of
Schaller~{et al.}  (1992) to fit both features well.  However, fits to the
upper main sequence lower metallicity isochrones can also be made with
appropriate extra reddenings, although the fit to fainter features in the
CMD worsens.  NGC~2214 lies in a region of $\sim$ 5 $\times $ $10^{19}$ H {\sc
i} atoms per $\rm cm^2$ (a column density), according to the Mathewson~\&
Ford (1983) map of the Magellanic Stream. Sauvage~\& Vigroux (1991) give
a relation of $E$($ B - V $) increasing by one magnitude for every 2.4~$\times
$ $10^{22}$ atoms per $\rm cm^2$ in the LMC. Reddening by this gas would
therefore not be detected.  Mould, Xystus~\& Da Costa (1993) give the
foreground reddening $E$($ B - V $) for the LMC as 0.07, in agreement with
the Galactic reddening maps of Burstein~\& Heiles (1982), which place
NGC~2214 between the 0.06 and 0.09 reddening contours.

The field red giant clump fell on the Swiss 1-Gyr isochrone, although a
continuous series of older clumps was present. However, good fits to the
lower sections of the subgiant branch required the lower metallicity of
0.004. The {\it Z}~= 0.001 isochrones were too blue, and could not fit this
feature well, whereas the 0.004 isochrones could also fit the slight
clumping of stars to the upper left of the giant clump, as well as the few
stars above the giant branch (i.e. as being on the asymptotic giant
branch). This would be in keeping with the general nature of the
age--metallicity relation of Da Costa (1991) for the LMC. The relative
densities along the giant clumps also suggest that there was a final burst
of field star formation $\sim$1-Gyr ago, after what appears to be continuous
formation as evidenced by the extended subgiant branch. Such a conclusion
is consistent with the results of Bertelli et al. (1992).

No isochrone could be fitted to the faint red stars given the adopted
distance modulus, although their distribution in the CMD resembled a giant
branch. The same feature can be seen in the field CMD. If this is a giant
branch, then it must belong to a population with a distance modulus greater
than that of the cluster. A program was used to assign an identification
number to each star, and could be used to plot the distribution of these
stars either on a CMD or on the CCD frame itself, allowing the physical
distributions of CMD features to be examined. The faint red stars were not
clumped, so another more distant cluster could not have been the cause of
the feature. The photometry of these stars was good.  None of them exceeded
the sharpness and $\chi$ limits of Bhatia~\& Piotto (1994), who explained
the older supergiant branch of Sagar~{et al.} (1991b) as being due to
photometric errors primarily caused by crowding in the central regions of
the cluster.  For `good' photometry, Bhatia~\& Piotto (1994) required
$\chi$ to be under 2 and the absolute value of the sharpness to be under
0.5. The faint red stars in the current CMDs had mean sharpness values of
0.95~$\pm$ 0.13 and 1.00~$\pm$ 0.15, and mean $\chi$ values of
$-0.01$~$\pm$ 0.05 and 0.00~$\pm$ 0.02, for the long-exposure {\it B} and
{\it V} frames respectively. These objects are therefore entirely
consistent with stellar images. It is highly unlikely that they represent
images of other objects, such as faint galaxies.

\begin{table}
\caption[Expected field star counts]{ { Expected field star counts.}
\label{table:scounts} The predicted star counts of Ratnatunga~\& Bahcall (1985)
for given regions of a CMD for the direction of the LMC are compared with
actual star counts from cluster CMDs obtained by this study. The region limits
are those given in Ratnatunga~\& Bahcall~(1985). The `O' rows give the observed
star counts, while the `E' rows contain the expected numbers.}
\vspace{-3mm}
\begin{center}
\scriptsize {
\begin {tabular}{@{}l||c|c|c|c|c|c@{}}
\hline
  & ($ B - V $) & \multicolumn{5}{c||}{$ V $ Magnitude Range} \\ \cline{3-7}
 & Colour & 13-15 & 15-17 & 17-19 & 19-21 & 21-23 \\
\hline
O  & 0.8-1.3 & 2 & 10 & 132 & 120 & 27 \\
E & 0.8-1.3 & 0.7 & 3.5 & 7.3 & 5.9 & 10 \\
\hline
O: & 1.3+ & 3 & 14 & 13 & 19 & 4 \\
E: & 1.3+ & 0.1 & 0.8 & 5.0 & 17.7 & 39.5 \\
\hline
\end{tabular}
}
\end{center}
\end{table}

Mateo~\& Hodge (1986) noted a similar `arm' in the remote LMC cluster NGC
1777, and considered it to be a subgiant branch at least 3-Gyr old, based
on Vandenberg's (1985) isochrones.  Bertelli~{et al.} (1992) considered that
similar stars in their CMDs were likely foreground objects, according to
the counts predicted by Ratnatunga~\& Bahcall (1985) for the direction of
the LMC. A comparison of the cluster CMD given in this study with the
predicted star counts given by the Bahcall~\& Soniera (1980, 1984) galaxy model
was made.  Observational counts exceeded the predicted ones in all but the
region {\it V} = 21 to 23 and ($ B - V $) greater than 1.3 where the
observational limit was met (see Table~\ref{table:scounts}).  The majority
of the enhancements were obviously due to features such as the giant
branch.  The region (V, $B - V $) = (21--23, 0.8--1.3) is to the red of the
field main sequence (MS) by more than any of the individual observational
errors of MS stars at these magnitudes, and it contains 3 times as many
stars as expected from the model, even though the count is not corrected
for incompleteness (see below) which will be significant at such faint
magnitudes.  The photometric errors of these stars do not overlap with
those of similarly faint MS stars. The region (19--21, 1.3+) is barely above
the expected value, and not statistically significant. However, when
counting incompleteness is considered, using the Lesser Completeness Factor
Method discussed below, some 26 stars are expected in this region. It
appears that foreground contamination or photometric uncertainty (such as
these being stars `scattered' off the main sequence) cannot account for
all the faint red stars, leaving their origin still unclear.


\subsection{Aperture photometry}

The {\sc iraf} {\sc imfort} program of Fischer~{et al.} (1992a) and
Fischer, Welch~\& Mateo (1992b; 1993) was slightly modified to handle the
AAT data and used to perform photometry, in a manner similar to Djorgovski
(1987), on the 30-s and 60-s {\it V} and {\it B} exposures of both NGC~2214
and the field region. The cluster frames were broken up into a series of
concentric elliptical annuli centred on the cluster centre. The same
central pixel coordinates were used for the field frames. Each annulus was
divided into eight sectors, and the intensity summed within each. The
median value was adopted as being representative of the sectors at the
weighted average radius, in an attempt to reduce the effect of bright
supergiants in the profile (Fischer~{et al.} 1993). The standard error of
the sectors is equal to the standard error of the mean multiplied by $
\sqrt { \pi / 2 } $, leading to the photometric uncertainty for the
annulus.

\begin{figure*}
\centerline
        {
        \psfig{figure=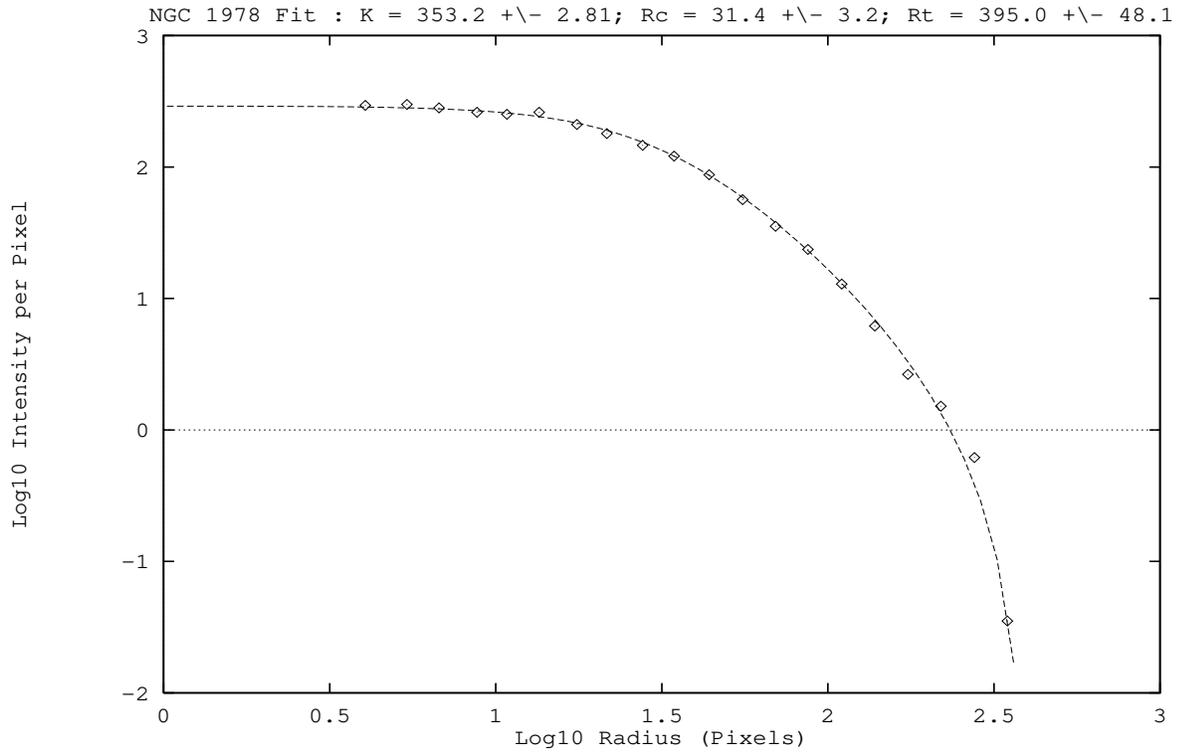,height=10.5cm,width=16cm,angle=-90}
        }
  { \caption[King model fit to NGC 1978]{ {  King model fit to NGC 1978.}
The logarthmic binned {\it V} profile for NGC 1978 is shown,
with the best-fitting King model.
\label{figure:atreides} } }
\end{figure*}

The program was used with the NGC 1978 Johnson {\it B}-band data of
Fischer~{et al.}  (1992b), as a check to see that it was performing
correctly. In this case an ellipticity of 0.3 was adopted.  A smooth
luminosity profile was derived, very similar to Fig.~10 of Fischer~{et al.}
(1992b), which was reassuring.  Tests using a circular aperture showed no
significant systematic differences from results using the elliptical
aperture, confirming the findings of Fischer~{et al.} (1992b), and lending
support to the contention of Elson~{et al.} (1987) that circular apertures
could be used on LMC clusters despite the (small on average) cluster
ellipticities.

King (1962) showed that three parameters were needed to describe the structure
of globular
clusters:
\begin{enumerate}
\item A core radius ($ r_{\rm c} $), specifying the size of a region of
constant density
near the cluster centre;
\item A tidal radius ($ r_{\rm t} $), being the tidal limit defined by the
gravity of the host galaxy;
\item A scaling factor $k$.
\end{enumerate}
The empirical formulation of King (1962) is given as:
\[
{ f = k \left(
\frac{1}{\sqrt{ 1 \: + \: ( \frac{r}{r_{\rm c}} )^2 } } \: - \:
\frac{1}{\sqrt{1 \: + \: ( \frac{r_{\rm t}}{r_{\rm c}} )^2 } } \right)^2 }
\]
where $f$ is the intensity per pixel (or surface density),
and $r$ the radius. King (1966) comments that his
later dynamical models agree closely with the
empirical curves. This expression was used as the
fitting equation for non-linear least-squares
optimizations.

King models were fitted to the {\it V}-band profile of NGC 1978, and the
{\it V} and {\it B} profiles of NGC~2214. NGC 1978 was fitted well by the
models. A typical result was $ r_{\rm c} \: = \: 12.2 \: \pm \: 1.4
$-arcsec and $r_{\rm t} \: = \: 154 \: \pm \: 5 $-arcsec (see
Fig.~\ref{figure:atreides}).  Slightly different starting parameters led
to different final values, with ranges of around 10-arcsec $<$ $ r_{\rm c} $ $
< $ 14-arcsec and 140-arcsec $<$ $ r_{\rm t} $ $<$ 170-arcsec, suggesting that
the minimum was rather flat. This was confirmed by a grid search, in which
$ r_{\rm c}$ and $ r_{\rm t}$ were fixed, leaving only $ k $ to be optimized.
A shallow, but fully enclosed, minimum was revealed.

The errors given are based on the photometric errors discussed above, and a
Hessian matrix (as in Bevington 1969). This matrix consists of the
second-order derivatives of $\chi^2$ against various parameters and
combinations of parameters about the optimum.  Mateo, Hodge~\&
Schommer(1986) noted that similar small errors in their $ \chi^2$
minimization King model fits were an illusion, and that many profiles
fitted the data nearly as well (see their Fig.~9). However, the derived
standard deviations are given below to indicate relative uncertainties
between parameters.  The core radius value for NGC 1978 agrees well with
the 11.2-arcsec estimate of Mateo (1987).

The NGC~2214 data were collected into equal-width log radius bins, with the
mean being taken for each populated bin, to avoid implicit weighting of the
model fit by more data points being in the outer regions. Fits to both the
binned and unbinned values required large values for the tidal radius.
Grid searches were also performed on the {\it B} and {\it V} NGC 2214
data. They revealed that, while the core radius was constrained, there was
only a lower boundary for the tidal radius.  The best {\it B} fit was for
radii $ r_{\rm c} \: = \: 9.9 \: \pm \: 2.4 $-arcsec and $ r_{\rm t} \: =
\: 296 \: \pm \: 67 $-arcsec (see Fig.~\ref{figure:atreides2}).  It was
repeatedly reached from various starting values, as long as $\rm r_t$ was
initially small.  The {\it V}-band result was similar in $ r_{\rm c} $,
being 10.6 $\pm$ 2.2-arcsec, and the tidal radius was 258 $\pm$
190-arcsec. The tidal radii were very dependent on the background values
adopted -- a variation of $ \sim$1 per cent could cause the radius to vary
by nearly a factor of 2.  The NGC~2214 radii agree well with the value of $
10.5 \: \pm \: 0.7 $-arcsec given by Elson (1991) for the core radius, and
the tidal radius range of 125 to 630-arcsec reported by Elson~{et al.}
(1987).

\begin{figure*}
\centerline
	{
	\psfig{figure=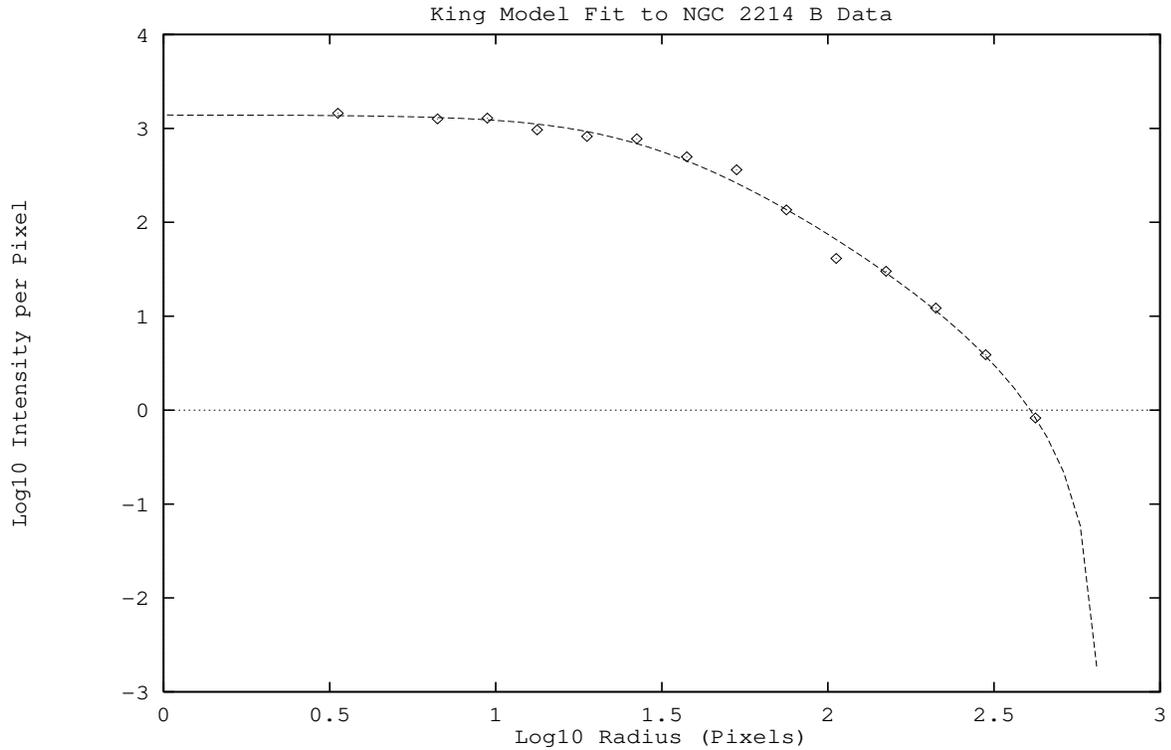,height=10.5cm,width=16cm,angle=-90}
	}
{
	\caption[King model fit to NGC~2214]
	{ {  King model fit to NGC~2214.}
	The log binned {\it B} profile for NGC~2214 is shown with  the best-fitting
King model. The
	gradient for NGC 2214
	is shallow and insufficient to provide good constraints on the tidal radius,
unlike
	NGC 1978.
	\label{figure:atreides2} }
}
\end{figure*}

Elson~{et al.} (1987) noted similar problems with constraining the tidal
radius in young LMC clusters, and commented that an unbound `halo' of stars
about the clusters might be responsible. Given that a King model fits the
intermediate-age cluster NGC 1978 well, but not the young cluster NGC 2214,
this conclusion is supported by the current study. Elson et al. (1987)
introduced a model intended to fit the profiles better: \[ { \mu ( r ) \: =
\: \mu_{0} \left ( 1 \: + \: \frac{r^2}{a^2} \right )^{\frac{- \gamma}{2}}
} \] which gave inconclusive results when it was used as the fitting
function to the current NGC~2214 data.  The form of this expression was
chosen purely for mathematical convenience by Elson~{et al.} (1987).  The $
\chi^2 $ for the {\it B}-band was $\sim$75 per cent better than that of the
best King model fit, but the {\it V}-band was $\sim$44 per cent worse,
making it unclear if the model fitted the data better.  The values of the
radius $a$ were 14.0 $\pm$ 0.2-arcsec and 16.6 $\pm$ 0.2-arcsec
respectively, which are larger than the value of 11-arcsec (no error
given) of Elson~{et al.}  (1987).  However, $\gamma$ values of 2.65 $\pm$
0.19 and 3.05 $\pm$ 0.22 were obtained for the two profiles. As commented
above, the standard deviation errors are an underestimate.  The first value
of $\gamma$ is in reasonable agreement with the value of 2.40 $\pm$ 0.24
derived by Elson~{et al.} (1987) for the cluster, while the second falls
into the range of $ \gamma $ covered by their young cluster sample. Elson
(1991) fitted such models to CCD observations of NGC 2214, but did not
present the $ \gamma $ values, commenting that they too fell into the same
range found by Elson~{et al.} (1987).

There is a potential problem with the aperture photometry program used in
this section, in that it simply checks to see if the {\em centre} of a
given pixel falls within an annulus. It does not consider how much of a
pixel may fall within the annulus, and scale the intensity
appropriately. Such a ragged edge to the annuli may become more important
with decreasing radius from the centre, in a region possibly critical for
fitting King (1966) models, although these values tend to be within the
core radius and so of essentially uniform intensity. This may be countered
by the observation above that elliptical and circular apertures resulted in
similar profiles. Despite these comments, the program can be used with no
reservations to search for colour gradients in the cluster, provided the
same parameters are used for the {\it B} and {\it V} frames, and so the
same pixels fall into the same apertures.  The field showed no colour
variation, nor any gradient in star counts. The cluster showed no
variation, beyond random scatter, at large radii, and then became
increasingly more red with decreasing radius as the ratio of cluster stars
to field stars increased. Finally, the colour levelled out within the main
body of the cluster, although the cluster is really too small for these
results to be definitive. Small-radii colours can be easily affected by
individual bright stars.

Aperture photometry was performed on the short-exposure images of the
cluster.  The aperture radius was 78-arcsec, giving sky-subtracted values
of {\it V} = 10.86 and $B - V$ = 0.36 mag. This is more red than
shallower aperture photometry such as that of van den Bergh~\& Hagen (1968)
and Elson~{et al.} (1987), which only measured the brighter population.


\subsection{Completeness}

%
%
\begin{table*}
\caption[Long exposure completeness cactors]{ { Long exposure completeness
factors} \label{table:kore_one} for the long-exposure {\it BV} field and
cluster
frames. Observational magnitudes are used (see text for reasoning). `Err' gives
the standard deviation error in the recovery rates.  }
\begin{minipage}{175mm}
\footnotesize{
\small{
\begin{center}
\begin {tabular}{@{}c|c||r|r|c|||r|r|c||r|r|c|c||r|r|c|c@{}}
\hline
\multicolumn{2}{||c||}{Range} & \multicolumn{3}{c||}{$V$ field} &
\multicolumn{3}{c||}{$B$ field}
	& \multicolumn{4}{c||}{$B$ cluster} & \multicolumn{4}{c||}{$V$ cluster} \\
\cline{1-2} \cline{3-5} \cline{6-8} \cline{9-12}  \cline{13-16}
Min & Mag & Out & In & \% & Out & In & \% & Out & In & \% & Err & Out & In & \%
& Err \\
\hline
14.0 & 14.5 & 529 & 550 & 96.2  & 887 & 900 & 98.6  &  3069 & 3100 & 99.0 & 1.8
& 2435 & 2450 & 99.4 & 2.0 \\
14.5 & 15.0 & 531 & 550 & 96.5  & 443 & 450 & 98.4  &  2220 & 2250 & 98.7 & 2.1
& 2225 & 2250 & 98.9 & 2.1 \\
15.0 & 15.5 & 526 & 550 & 95.6  & 442 & 450 & 98.2  &  3485 & 3550 & 98.2 & 1.7
& 2457 & 2500 & 98.3 & 2.0 \\
15.5 & 16.0 & 537 & 550 & 97.6  & 444 & 450 & 98.7  &  1717 & 1750 & 98.1 & 2.4
& 2168 & 2200 & 98.6 & 2.1 \\
16.0 & 16.5 & 535 & 550 & 97.3  & 441 & 450 & 98.0  &  3052 & 3100 & 98.4 & 1.8
& 3434 & 3500 & 98.1 & 1.7 \\
16.5 & 17.0 & 539 & 550 & 98.0  & 448 & 450 & 99.6  &  2591 & 2650 & 97.8 & 1.9
& 2354 & 2400 & 98.1 & 2.0 \\
17.0 & 17.5 & 534 & 550 & 97.1  & 443 & 450 & 98.4  &  3169 & 3250 & 97.5 & 1.5
& 3573 & 3650 & 97.9 & 1.7 \\
17.5 & 18.0 & 535 & 550 & 97.3  & 437 & 450 & 97.1  &  2038 & 2100 & 97.0 & 2.2
& 3066 & 3150 & 97.3 & 1.8 \\
18.0 & 18.5 & 543 & 550 & 98.7  & 442 & 450 & 98.2  &  3801 & 3950 & 96.2 & 1.6
& 2832 & 2900 & 97.7 & 1.9 \\
18.5 & 19.0 & 541 & 550 & 98.4  & 443 & 450 & 98.4  &  2733 & 2850 & 95.9 & 1.9
& 2649 & 2750 & 96.3 & 1.9 \\
19.0 & 19.5 & 537 & 550 & 97.6  & 441 & 450 & 98.0  &  3222 & 3350 & 96.2 & 1.7
& 3151 & 3250 & 96.9 & 1.7 \\
19.5 & 20.0 & 533 & 550 & 96.9  & 433 & 450 & 96.2  &  2489 & 2650 & 93.9 & 1.9
& 4633 & 4800 & 96.5 & 1.4 \\
20.0 & 20.5 & 529 & 550 & 96.2  & 429 & 450 & 95.3  &  2910 & 3200 & 90.9 & 1.8
& 2268 & 2450 & 92.6 & 2.0 \\
20.5 & 21.0 & 514 & 550 & 93.5  & 420 & 450 & 93.3  &  2921 & 3350 & 87.2 & 1.7
& 1442 & 1600 & 90.1 & 2.5 \\
21.0 & 21.5 & 543 & 600 & 90.5  & 409 & 450 & 90.9  &  2921 & 3500 & 83.5 & 1.7
& 1674 & 2000 & 83.7 & 2.2 \\
21.5 & 22.0 & 515 & 600 & 85.8  & 423 & 500 & 84.6  &  1986 & 2500 & 79.4 & 2.0
& 1633 & 2100 & 77.8 & 2.2 \\
22.0 & 22.5 & 538 & 700 & 76.9 & 1060 & 1350 & 78.5 &  3231 & 4550 & 71.0 & 1.5
& 1781 & 2450 & 72.7 & 2.0 \\
22.5 & 23.0 & 727 & 1200 & 60.6 & 422 & 650 & 64.9  &  1990 & 3200 & 62.2 & 1.8
& 771  & 1250 & 61.7 & 2.8 \\
23.0 & 23.5 & 526 & 1550 & 33.9 & 400 & 850 & 47.0  &  1003 & 2150 & 46.7 & 2.2
& 1627 & 4000 & 40.7 & 1.5 \\
23.5 & 24.0 & 236 & 1500 & 15.7 & 291 & 1050 & 27.7 &  536  & 2250 & 23.8 & 2.1
& 566  & 2400 & 23.6 & 2.0 \\
24.0 & 24.5 & 61  & 550  & 11.1 &  60 & 450 & 13.3 & - & - & - & - & - & - & -
& - \\
\hline
\multicolumn{2}{||c|}{Total:} & 9786 & 10800 & - & 9658 & 12050 & - & 51084 &
59520 & - & - & 47639 & 54050 & - & - \\
\hline
\end{tabular}
\end{center}
}
}
\end{minipage}
\end{table*}

\noindent
The brighter stars in a frame are almost certainly all recovered by the
crowded field reduction software. However, certainly not all of the
faintest magnitude stars are identified and recovered, with the recovery
rate decreasing with increasing magnitude.  Problems affecting image
recovery include the stars
\begin{enumerate}
\item simply being so faint that their identification is adversely affected by
stochastic variations of the background;
\item being too close to a comparably bright star, leading to a blended
elliptical object
which will be rejected as not being a star;
\item being lost in the profile of a much brighter star.
\end{enumerate}
A luminosity function that is not `corrected' for these effects will have a
more shallow gradient, and with increasing magnitude increasingly
underestimate the actual function.  Empirical methods of estimating these
`completeness factors' generally centre around the addition of artificial
stars, or scaled versions of the PSFs derived from selected images, into
the CCD frames. These frames are then reduced in an identical manner to the
original frame. The efficiency with which these false stars are recovered
is taken to estimate the completeness factors, which are then used to
correct the observed main-sequence star counts to the values expected if
the recovery rate had been 100 per cent.  For a given magnitude interval $ i $
the
completeness factor is given as: \[ \Lambda_{i} = \frac{ n_{i,\:{\rm
recovered}}}{ n_{i,\:{\rm added}}} {\rm \: . } \] On a single frame the
completeness correction can easily be determined since the variation of the
completeness factors with magnitude and crowding can be empirically
determined. However, it is not possible to separate main-sequence stars out
of the mass of stars. Two colours, such as {\it B} and {\it V}, are
required for their identification.  The resulting problem of having to
match the star image in both colours leads to a further
incompleteness. Mateo (1988) and Mateo~\& Hodge (1986) did not account for
this last point, although mention was made of it, and simply considered
that the completeness correction for a point $ { ( V_i , \: B_i \: - \: V_i
) } $ would be given by: \[ \Lambda ( V_i ) \: {\rm \times } \: \Lambda (
B_i ) \: .  \] Sagar~\& Richtler (1991) argued that the two frames were
not independent and that the multiplicative assumption of Mateo (1988)
could not be justified. Instead, as the spatial distribution of stars in
the frames is the same and the magnitude distribution is slightly modified,
the completeness at a given point in the CMD would be mainly controlled by
the lesser of the two completeness factors.  There has been no test of the
ability of these techniques to recover a known luminosity function.

\begin{table}
\caption[Short exposure field frames completeness factors] { { Short exposure
field frames completeness factors} \label{table:kore_three} are given for the
observational magnitudes.}
\scriptsize{
\begin{center}
\begin{tabular}{||c|c|r|r|c|r|r|c||}
\hline
\multicolumn{2}{||c|}{Range} & \multicolumn{3}{c|}{$V$ Field} &
\multicolumn{3}{c||}{$B$ Field} \\ \cline{1-8}
Min & Mag & Out & In & \% & Out & In & \% \\
\hline
14.0 & 14.5 &  711 & 720  & 98.8 & 1155 & 1170 & 98.7 \\
14.5 & 15.0 &  709 & 720  & 98.5 & 1157 & 1170 & 98.9 \\
15.0 & 15.5 &  707 & 720  & 98.2 & 1923 & 1950 & 98.6 \\
15.5 & 16.0 &  705 & 720  & 97.9 & 1932 & 1950 & 99.1 \\
16.0 & 16.5 &  771 & 780  & 98.9 & 1890 & 1900 & 99.5 \\
16.5 & 17.0 &  590 & 600  & 98.3 & 1937 & 1950 & 99.3 \\
17.0 & 17.5 &  769 & 780  & 98.6 & 1223 & 1230 & 99.4 \\
17.5 & 18.0 &  766 & 780  & 98.2 & 509 & 510 & 99.8  \\
18.0 & 18.5 &  767 & 780  & 98.3 & 1217 & 1230 & 98.9 \\
18.5 & 19.0 &  762 & 780  & 97.7 & 768 & 780 & 98.5 \\
19.0 & 19.5 &  811 & 840  & 96.6 & 1146 & 1170 & 97.9 \\
19.5 & 20.0 &  799 & 840  & 95.1 & 1184 & 1230 & 96.3 \\
20.0 & 20.5 & 2294 & 2400 & 95.6 & 1180 & 1230 & 95.9 \\
20.5 & 21.0 & 1446 & 1560 & 92.7 & 414 & 450 & 92.0 \\
21.0 & 21.5 &  732 & 840  & 87.1 & 1148 & 1450 & 79.2 \\
21.5 & 22.0 & 1431 & 1860 & 76.9 & 1112 & 2790 & 39.9 \\
22.0 & 22.5 & 1416 & 2700 & 52.4 & 107 & 1050 & 10.2 \\
22.5 & 23.0 &  322 & 1800 & 17.9 & - & - & - \\
\hline
\multicolumn{2}{||c|}{Total:} & 16448 & 20220 & - & 20002 & 23210 & - \\
\hline
\end{tabular}
\end{center}
}
\end{table}

In order to estimate the completeness factors for our data, an {\sc iraf}
script was written which placed a small number of artificial stars into a
frame.  50 stars was the maximum number of stars added at each iteration by
the script (being typically $ \sim5$ per cent of the detected stars), which
meant that the frame crowding and luminosity function would not be greatly
affected by the introduction of the artificial stars.  A user-selected
magnitude range was divided up into bins, typically 0.5 mag wide.  `Random'
magnitudes and ($x$, $y$) positions were generated for stars in each bin,
using the `Minimal Standard' pseudo-random number generator of Park~\&
Miller (1978). Press ~{et al.} (1992, p.\ 279) note that the period of this
generator is $ 2^{31} \: - \: 1 $.

Several sets of false stars were added to the frame (in different
iterations) and reduced until a user-set limit of recovered false stars was
met.  This limit was set high since the completeness factors vary with
crowding. In a frame of a LMC star cluster, the degree of crowding
naturally varies across the frame. Mateo (1988) divided frames into three
rings centred on the cluster, surrounded by field, and formed a mean
luminosity function from these rings.  Rather than adopt this technique,
and its assumptions (see Mateo 1988), we chose to derive a mean by placing
many stars into the frames.

\begin{figure*}
\centerline
	{
	\psfig{figure=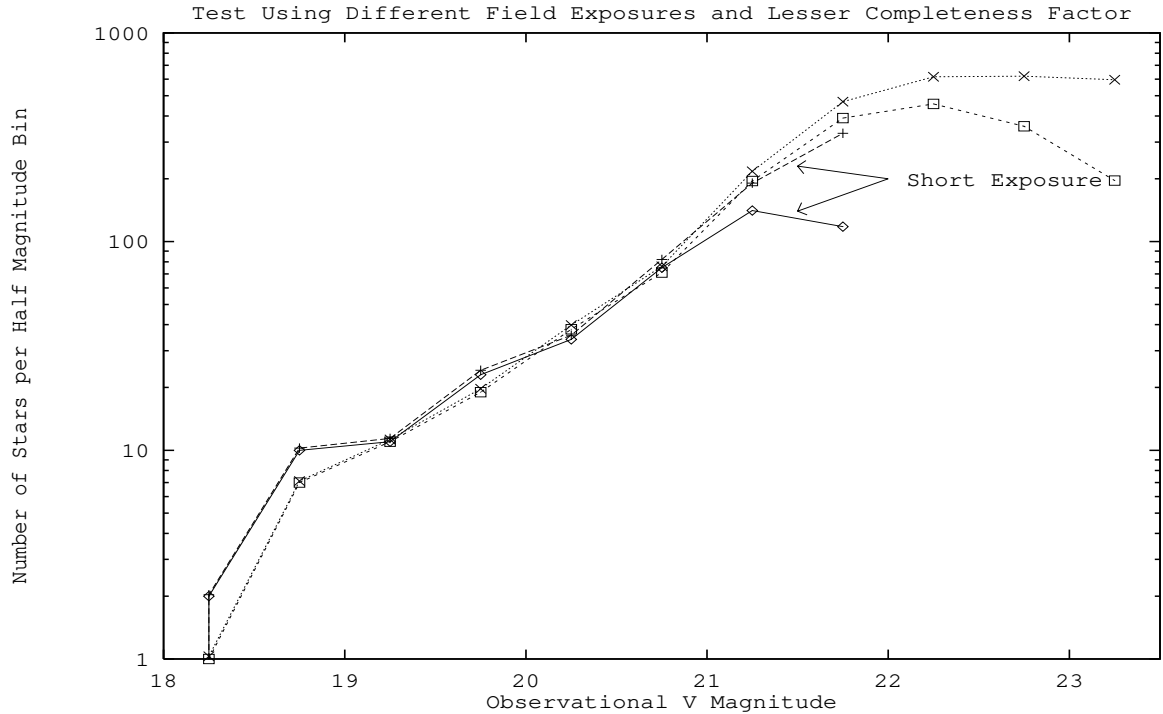,height=10cm,width=16cm,angle=-90}
	}
\caption[Observational {\it V}-band luminosity functions]{{ Observational {\it
V}-band luminosity functions.} The long- and short-exposure {\it V}-band
luminosity functions, with no attempt to remove non-main-sequence stars, are
plotted both with and without the appropriate completeness corrections applied.
The uncorrected (raw) data are given by $\Diamond$ and $\Box$ for the short and
long exposures respectively, while the corrected values are indicated by + and
$\times$ symbols.  \label{figure:maimai2}}
\end{figure*}

Several {\sc iraf} scripts were written to test techniques for estimating
completeness factors. The tests involved the creation of artificial field
frames with known luminosity functions. The completeness techniques were
then used to estimate these functions. These estimates could be compared
with the actual function.  Details of these tests can be found in Banks
(1994). The product method of Mateo (1988) was found increasingly to
overestimate the completeness correction as magnitude increased. The Lesser
Ratio method recovered the actual luminosity function better, with a mean
error of $\sim3$ per cent, although near the observational limit of the
frames the technique underestimated the function. The technique assumed
that all the false stars recovered on the frame producing the lower
completeness factor were all recovered on the second frame. Towards fainter
magnitudes, recovery rates drop and so this assumption cannot hold. Another
computer program was written to test the effect of this assumption. Stars
{\em with} colour were placed into both the {\it B} and {\it V} frames, and
recovered in the same manner as the `real stars' were. This dual-frame
method estimated the recovery rates better than the other techniques,
except when the factor fell below 50 per cent, which Stetson (1991) defines
as the limiting magnitude of a CCD frame. However, the technique is {\em
extremely} computer-intensive, making use of it prohibitive if it is to be
used over the entire magnitude range of the frames.  Fortunately, the
matching problem will only affect completeness estimates near the faint
limit of the images, allowing the simpler Lesser Ratio method to be used
for the bulk of the calculations.

The results of the completeness factor trials on the
long-exposure frames are given in Table~\ref{table:kore_one}.
More artificial stars were added to the cluster frames in an attempt to
account for the variation in crowding across the frames.

The completeness factors for the short-exposure field frames are in
Table~\ref{table:kore_three}.  These values were used in conjunction with
the Lesser Ratio method to `correct' the short-exposure {\it V} luminosity
function.  Similarly, values from Table~\ref{table:kore_one} were used to
correct the long-exposure {\it V} function. This was to test whether the
same `real' function would be recovered from the two exposures. It was
found to be true in general, except for the brightest and faintest
magnitudes (see Fig.~\ref{figure:maimai2}).  The difference at faint
magnitudes is likely to be due to the matching problem mentioned above
(which led to the dual-frame method), which \label{disc} will become more
apparent near the limiting magnitude. Use of the dual-frame completeness
script seems to be warranted when the individual frame completeness factors
fall below $\sim$75 per cent. The flattening of the long-exposure {\it V}
luminosity function shown in Fig.~\ref{figure:maimai2} lies in such a
range.

The difference at bright magnitudes is due to saturation of stars in the longer
exposures. The
position of a star within the cluster will affect its recovery, depending on
the
star's magnitude. Bright stars, no matter where they are in a frame, will be
saturated in a suitably long exposure. However, there will be a magnitude range
where location does matter. If such a star is in a relatively uncrowded region
of the frame, it will not be saturated. However, if it is in a crowded region
then its light, combined with the fainter stars that it is covering, will cause
the
stellar image to saturate.


\subsection{Field star subtraction}

Field stars need to be statistically subtracted from the luminosity
function of the cluster. Cluster membership itself is hard to assess in the
LMC. Even if radial velocities are available for all the stars, Freeman,
Illingworth~\& Oemler (1983) showed young clusters to have disc dynamics,
making such data rather uninformative about membership, only excluding
Galactic stars.  Flower~{et al.}  (1980) and Olszewski (1984) subtracted a
star from the cluster CMD for every star within the same given region in
the field CMD. This technique ignored completeness factors being different
between the field and the cluster.  Mateo~\& Hodge (1986) adjusted the
number of stars in both the field and the cluster by their completeness
factors and also by the ratio of the areas of the field and cluster
regions.

A computer program was written to subtract field stars from the cluster
frame listing.  Both the {\it B} and {\it V} lists were based on the
long-exposure frames. The observational CMD could be divided up into a
({\it B}, {\it V}) or a ({\it V}, $ B - V $)} grid, or into circular
regions about each star in the cluster frame. Regardless of which of the
three methods was used to bound a region in the CMD, the numbers of stars
within this region in the field and cluster CMDs were counted. The
completeness factors for the region centre in both the field and cluster
CMDs were used to adjust the star counts.  The selected regions therefore
should not to be too large, otherwise the completeness factors would vary
within the region itself.  The completeness corrected cluster to field
ratio was taken as the probability that a given star (within the region) in
the cluster star list was actually a cluster member. If a grid was being
used then a random number was generated for each cluster list member within
the region. If the random number fell above the probability given by the
corrected ratio, the star was considered to be a cluster member, otherwise
it was removed from the list. In the case of the circular regions, the
probability was applied only to the central star, as all the stars in the
cluster frame would be checked in turn. It should be noted that the cluster
star list was not altered during the execution of the program.  Artificial
star and completeness factor lists were used to test the software, which
performed as expected. Given the result of the completeness tests, the {
lesser} of the appropriate {\it B} and {\it V} completeness factors was
used in the analysis of NGC~2214.

\subsection{Luminosity functions}

The field-star-subtracted star lists need to be corrected for counting
incompleteness, before a luminosity function can be estimated.  The field star
subtraction discussed above only compared the cluster and field ratios in order
to calculate a probability that a given star in the cluster CMD was a field
member. Mateo~(1988) commented that the calibration of the false star
magnitudes
was problematic as transformation equations explicitly involve a colour term,
and so assumed that all his false stars had a ($ B - V $) colour of 0.5 mag.

To overcome this problem a bin distribution was selected for the {\it V} and
{\it B} luminosity functions.  The standardized magnitudes of each star were in
turn transferred back to the observational system using the transformation
equations given above. The relevant {\it B} and {\it V} completeness factors
were then referenced, and the appropriate completeness ratio (Mateo, Lesser, or
Greater) chosen. The standardized magnitudes were used to determine which
luminosity function bins should be altered.  The appropriate bin counts were
then incremented by the inverse of the completeness ratio. This technique
avoided the problem of standardizing the false star magnitudes. The program
output results at every step of the process during testing. These values were
compared with manual calculations, and found to be correct.

In addition, only main-sequence stars should be included in a luminosity
function. Following Mateo (1988), a line was arbitrarily drawn between
main-sequence and evolved stars, with the evolved stars being
discarded. The importance of using two-colour photometry to define
main-sequence stars belonging to a cluster can be illustrated by the Da
Costa (1982) luminosity function for the Galactic globular cluster 47 Tuc.
A series of single-colour plates was used, producing a relatively steep
luminosity function. Later {\it BV} CCD photometry of the cluster by
Harris~\& Hesser (1985) revealed that there was severe contamination of the
cluster photometry by faint SMC stars. Removal of these stars flattened the
luminosity function to the point where 47 Tuc appears to have one of the
flattest luminosity functions amongst the globular clusters.

Several `subtractions' were performed with different seeds for the
pseudo-random number generator, producing similar results. The individual
search option was used, with a search radius of 0.282-mag. This corresponds
to the same area as that of a 0.5 square magnitude box, or roughly 1.4
times the area of a given combined completeness bin. The intention was to
use a search area large enough to collect a reasonable number of stars to
avoid low-number statistics, yet not so large that the completeness
corrections varied substantially with the region. Given that completeness
factors were calculated for 0.5-mag bins in each frame, it was an arbitrary
decision to use a region slightly larger than a completeness `diamond' (see
Fig.~10 in Mateo 1988) in order to satisfy these two requirements.

\subsection{Mass functions}

The luminosity functions were converted into mass functions using the best
fitting Swiss isochrone discussed above ($Z$~=~0.020 with a logarithm age
of 8.0) to derive a mass--luminosity function for the cluster. The
evolution in luminosity of a star still on the main-sequence can be quite
substantial. The change in luminosity between the zero age main-sequence
and the turnoff of a star can correspond to between 0.5 and 1.6 mag,
depending on the mass of a star (Mateo 1988).  We note that the global mass
function slope does not sensitively depend on the evolutionary models used
(see Mateo (1988) and Sagar~\& Richtler~(1991) who compare mass functions
derived using classical and overshooting stellar models).  As Mateo~(1993)
points out, the mass range being studied changes, but the gradient remains
effectively the same with the greatest variation occuring below
$Z$~=~0.004.

\begin{figure*}
\centerline
	{
	\psfig{figure=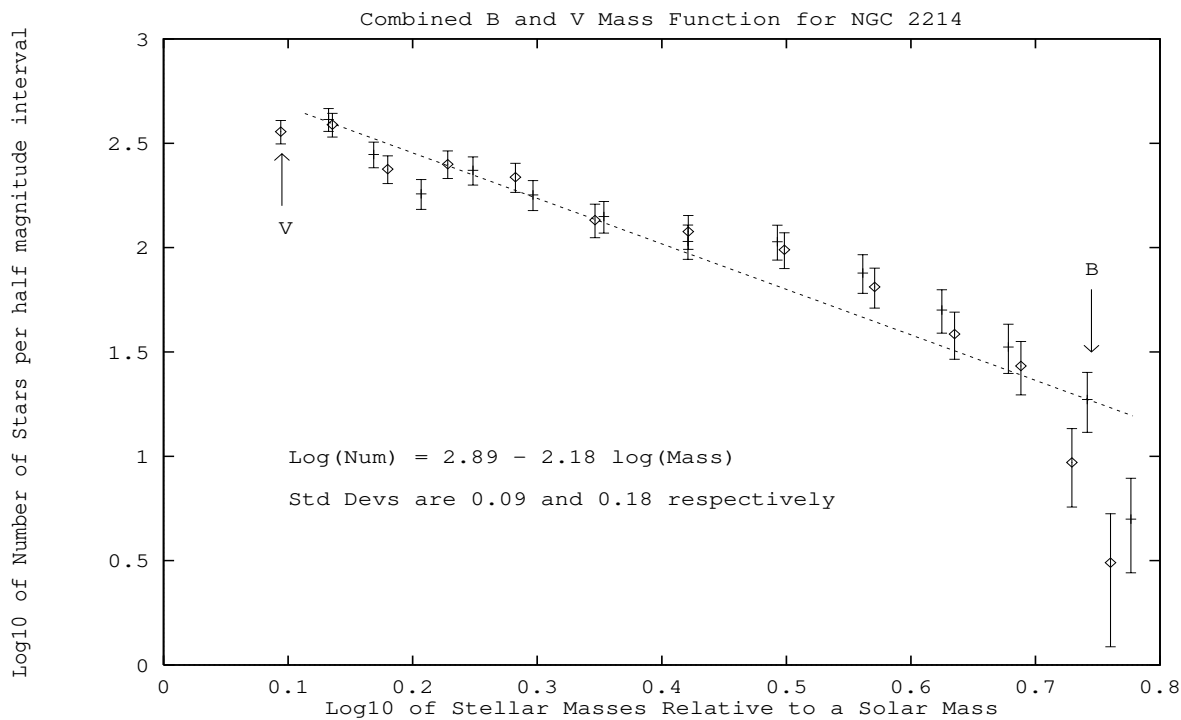,height=10cm,width=16cm,angle=-90}
	}
{ \caption[Mass function for NGC~2214 based on the long-exposure frames]
{{ Mass function for NGC~2214 based on the long-exposure frames.}
Values calculated from {\it B} magnitudes are given as horizontal dashes, while
the {\it V}-based ones are marked by diamonds. The dashed line is the linear
least-squares fit to
the {\it B}-data, and its parameter values are given.
\label{figure:massfn} } }
\end{figure*}

\begin{figure*}
\centerline
	{
	\psfig{figure=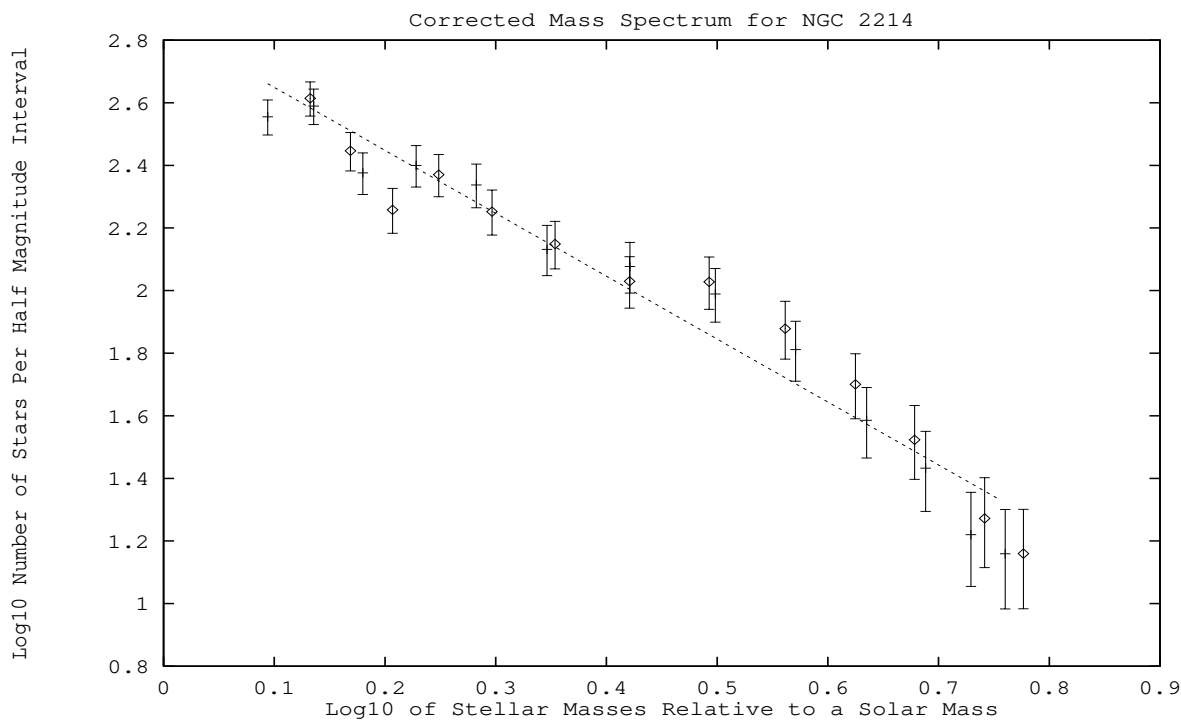,height=10cm,width=16cm,angle=-90}
	}
{ \caption[Mass function for NGC~2214]
{{  Mass function for NGC~2214.} The same symbols are used as in
Fig.~\ref{figure:massfn}, but the
high-mass bins are based on the short-exposure frames. The line of best fit
given in the previous
figure is shown, to ease comparison between the diagrams, and to demonstrate
just how uncertain the
gradient is.
\label{figure:newfn} } }
\end{figure*}

The resulting mass functions derived from the long-exposure frames are
given in Fig.~\ref{figure:massfn}. Functions were calculated from both the
{\it B} and {\it V} luminosity functions, and are displayed. The errors
shown are the Poisson errors in the number of actually retrieved stars
(i.e.  before the star numbers were corrected for completeness) combined
with uncertainty in the completeness correction.  The error in the
completeness factors themselves is hard to quantify. Poisson errors were
assumed for them (as in Table~\ref{table:kore_one}).  Their inclusion could
be quite involved -- as the stars falling into a given bin could belong to
subgroups each with a different completeness correction.  In order to have
an upper estimate, the largest of the completeness correction errors was
used for the entire group.  Linear least-squares fitting was performed on
the functions.  Using the expression for the IMF introduced in Section~1,
$x$ values of 1.18 $\pm$ 0.18, 1.47 $\pm$ 0.21, and 1.32 $\pm$ 0.14 were
derived for the {\it B}, {\it V}, and combined {\it BV} data.  These
gradients are in reasonable agreement with Sagar~\& Richtler (1991), who
calculated $ x $ values of 1.1 $\pm$ 0.3, 2.2 $\pm$ 0.3, 1.4 $\pm$ 0.5, and
1.3 $\pm$ 0.3 for NGC~2214, using the lesser of the {\it BV} completeness
factors and different stellar evolution models.

Like the results of Sagar~\& Richtler (1991), the mass function for
NGC~2214 appears to have a steep gradient at the higher masses. They made
no comment about this feature, and applied a single straight line.  The
steep decline in the two most massive {\it V} bins, and the most massive
{\it B} bin, is an artefact of the reduction process. While an isolated
star with a magnitude that fell into one of these bins would not be
saturated on the real frame, it would be if the star were in a region with
an enhanced background level, e.g. the cluster centre.  Unfortunately, no
clipping level corresponding to that used in the reduction of actual
observations was set for the artificial star trials. The artificial stars
were therefore not rejected even if they were in the cluster centre,
leading to an overestimate of the completeness factor on these, the
brightest of the bins.  It is possible that the same effect was present in
the artificial star trials of Sagar~\& Richtler (1991).

The short-exposure frames were examined for the number of stars that fell
into these bins. The numbers were substantially larger, and brought the
mass spectrum values of the bins up (see Fig.~\ref{figure:newfn}).  There
is still some hint that two lines could be fitted to the high- and low-mass
stars. An $x$ value of 2 fits the $\geq \rm 3 M_{\sun}$ stars well.  Below
$\sim3 \rm M_{\sun}$ a shallower gradient appears reasonable, corresponding
to an $x$ value of 0.8. The combined cluster mass function of Mateo~(1988)
was over effectively the same range as that of this study. He suggested
that the low-mass end (${\rm log} \: M/ \rm M_{\sun}$ $<$ 0.45) was
slightly steeper. We have found that the incompleteness technique of Mateo
(1988) under-estimates recovery rates, with the discrepancy increasing as
the completeness rates decline.  Therefore our gradients are substantially
less than those of Mateo (see Sagar~\& Richtler (1991) who applied the same
completeness technique corrections to NGC~1711, which was the only cluster
in common with Mateo (1988), and derived a similar gradient to that found
in the present study). The steep gradients of Mateo are not confirmed; nor
are the very shallow gradients of Elson~{et al.} (1989).

We considered the possibility that the change in gradient was due to
incorrect estimation of the recovery rates at faint magnitudes.  The
dual-frame completeness script was therefore used on the cluster and field
frames. The bins were 0.5~mag wide over the observational {\it V} magnitude
range 20--23, and were centred on the main-sequence with a ($ B - V$) width
of 0.30. These bins were selected as per the discussion at the end of
Section~3.8.  In the case of the current data, no advantage was gained.
The estimated completeness factors were within 1.2 $\pm$ 0.7 per cent of
those estimated by the lesser completeness factor method. This result
raises the question of whether more complicated dual-frame completeness
techniques, employing a variant of the matrix method proposed by Drukier~et
al. (1988) to measure the effect of the bin migration discussed above,
would be worthwhile in practice.

Linear least-squares fits to the {\it B}, {\it V}, and combined {\it B} and
{\it V} mass spectra (of all the data points) resulted in $x$ values of
0.96~$\pm$ 0.14, 1.06~$\pm$ 0.13, and 1.01~$\pm$ 0.09 respectively. While
this is shallower than the Sagar~\& Richtler~(1991) values for NGC~2214, we
note that we can recover similar values before we correct the saturation
problem discussed above, and that the new values are in reasonable
agreement with the average value of $x$ = 1.1 of Sagar~\& Richtler~(1991)
for their five LMC clusters overall for the mass range 1.9--6.3~$\rm
M_{\sun}$.

Sagar~\& Richtler~(1991) investigated the effect that binary stars have on
mass functions, finding that the slope of the actual mass function
determined the effect itself.  Steep mass functions are weighted towards
low-mass stars.  Therefore since the secondary component of a binary is
likely to be less massive than the primary, and so causing little change in
the luminosity of the system.  Alternatively, flat distributions will
produce binaries with mass ratios closer to unity, dramatically affecting
the system luminosity (Mateo 1993). Sagar~\& Richtler~(1991) found that
mass functions with gradients of 2.5 or steeper were essentially unaffected
by binaries, while a mass function of slope 1.5 could be lowered by 0.4 in
the extreme case of every star being binary. While the actual proportion of
binary stars in LMC clusters is unknown, the work of Sagar~\&
Richtler~(1991) indicates that the gradients derived by studies similar to
this one are underestimates of the actual gradient.

Mateo (1993) and Banks (1994) review results of mass function studies of
Magellanic Cloud clusters. They note that these estimates are all similar,
despite being for quite different mass ranges and metallicities --
suggesting the possibility of a global IMF in the Clouds. There is poor
agreement between mass function estimate studies of the solar neighbourhood
(see Banks 1994), emphasizing the difficult nature of this work (see Scalo
1986 for a full discussion of the problems and assumptions involved).  It
is unclear whether the IMF of young Magellanic Clusters (MC) and
Associations is different from Galactic values, and no definite conclusions
can be made on the universality of the IMF, based on these results, at
present. It has been suggested that the IMF will flatten with decreasing
metallicity (see e.g. Terlevich~\& Melnick 1985; Piotto 1991), and it could
be argued that the MC IMFs are flatter -- although it should be noted that
isochrones of solar neighbourhood metallicity provided the best fit to the
NGC~2214 CMD, and there does not appear to be any difference between
Galactic halo and disc field star IMFs over the small mass range of 0.3 to
0.8~$\rm M_{\sun}$ (Scalo 1986).  Further study is required, including the
estimation of the mass functions of LMC clusters with substantially
different metallicities.

\section{Conclusions}

We have found no evidence supporting the contention that NGC~2214 is a
merging cluster. Models have been fitted to radial profiles of the cluster,
and support the contention of Elson~et al.\ (1987) that an unbound halo of
stars causes poor fits by King models. Techniques for evaluating
completeness methods have been tested, with the method of Mateo~(1988)
being shown to underestimate completeness factors. The best techniques were
employed on the AAT data, leading to an estimate of the cluster's mass
function, with an $x$ value of $\sim1$, which is in good agreement with
other studies of young Magellanic Cloud clusters. IMF estimates for
Galactic regions are unreliable, which makes any conclusions about the
universality of the IMF indistinct. However, there are no {\em
substantial}\/ differences between IMFs derived for the Magellanic Clouds
and for our Galaxy, and it is likely that star formation in these three
galaxies can be described by a `universal' IMF, at least over the mass
interval $\sim1$ to $\sim10$~$\rm M_{\sun}$.

\section*{Acknowledgments}

The authors are grateful to Dr G.\ Da Costa and the AAO for service
observing by the Anglo-Australian Telescope, Acorn NZ for the loan of an
R260, Dr D.\ Schaerer for isochrone tables, Dr Phillipe Fischer for his
aperture photometry program, Dr J. Scalo for offprints, the Institute of
Statistics and Operations Research (Victoria University of Wellington) for
extra computing facilities, Dr M.\ Mateo for his helpful comments as
Referee for this paper, and the Foundation for Research, Science and
Technology of New Zealand for partial funding of this project in
conjunction with the VUW Internal Research Grant Committee.  TB
acknowledges partial support during this study by the inaugural Royal
Society of New Zealand R.H.T.\ Bates Postgraduate Scholarship. {\sc iraf}
is courtesy of the National Optical Astronomical Observatories, which are
operated by the Association of Universities for Research in Astronomy,
Inc., under cooperative agreement with the US National Science
Foundation. The Space Telescope Science Data Analysis System ({\sc stsdas})
is courtesy of the Space Telescope Science Institute, Baltimore.


\label{lastpage}
\bsp

\end{document}